\documentclass[journal]{IEEEtran}
\ifx\pdfoutput\undefined
\bibliography{mybib}
\usepackage{graphicx}
\else
\usepackage[pdftex]{graphicx}
\fi
\usepackage{url}
\usepackage{array}
\usepackage{stfloats}
\usepackage{amssymb}
\usepackage{amsmath}
\usepackage{amsthm}
\usepackage{cite}
\usepackage{enumerate}
\usepackage{epstopdf}
\usepackage{color}

\DeclareGraphicsExtensions{.pdf,.png,.jpg}
\begin{document}
 \title{Energy-Efficient Symbol-Level Precoding in Multiuser MISO Based on Relaxed Detection Region }
 \author{
 \IEEEauthorblockN{ Maha~Alodeh,~\IEEEmembership{Student~Member, IEEE}, Symeon Chatzinotas, \IEEEmembership{Senior~Member,~IEEE,}
 Bj\"{o}rn Ottersten, \IEEEmembership{Fellow Member,~IEEE}\thanks{Maha Alodeh, Symeon Chantzinotas and  Bj\"{o}rn Ottersten are
with Interdisciplinary Centre for Security Reliability and Trust (SnT) at the University
of Luxembourg, Luxembourg. E-mails:\{ maha.alodeh@uni.lu, symeon.chatzinotas
@uni.lu, and bjorn.ottersten@uni.lu\}. \newline
This work is supported by Fond National de la Recherche Luxembourg (FNR)
projects,
project Smart Resource Allocation for Satellite Cognitive Radio (SRAT-SCR)  ID:4919957 and Spectrum Management and Interference Mitigation in Cognitive Radio Satellite Networks SeMiGod.}}\\
  
}
 
 \maketitle
 \IEEEpeerreviewmaketitle
 \begin{abstract}
\boldmath This paper addresses the problem  of exploiting interference among simultaneous multiuser transmissions in the downlink of multiple-antenna systems. Using symbol-level precoding, a new approach towards addressing the multiuser interference is discussed through jointly utilizing the channel state information (CSI) and data information (DI). The interference among the data streams
is transformed under certain conditions to a useful signal that can improve the signal-to-interference noise ratio (SINR) of the downlink transmissions and as a result the system's energy efficiency. In this context, new constructive interference precoding techniques that tackle the transmit power minimization (min power) with individual SINR constraints at each user's receiver have been proposed. In this paper, we generalize the CI precoding design under the assumption that the received MPSK symbol can reside in a relaxed region in order to be correctly detected. Moreover, a weighted maximization of the minimum SNR among all users is studied taking into account the relaxed detection region. Symbol error rate analysis (SER) for the proposed precoding is discussed to characterize the tradeoff between transmit power reduction and SER increase due to the relaxation. Based on this tradeoff, the energy efficiency performance of the proposed technique is analyzed. Finally, extensive numerical results
show that the proposed schemes outperform other state-of-the-art techniques.\\

\end{abstract}

\vspace{-0.1cm}
\section{Introduction}
%\vspace{-0.15cm}

Spatial division multiple access (SDMA) exploits the multiple antennas at the communication terminals to serve multiple users simultaneously \cite{roy}. Utilizing the same time and frequency dimensions, interference is one of the crucial factors that hampers its implementation. Multiple antennas introduces spatial degrees of freedom providing the separation of users and limiting the harmful effects of interference thereby allowing spatial multiplexing \cite{roy}-\cite{haardt}.

The applications of SDMA, in which a transmitter equipped with multiple antennas
aims to communicate with multiple receivers,
vary according to the requested
service. The first service type is known as a broadcast in which a transmitter has a common message to be sent to multiple receivers. In physical layer research,  this service has been studied under the term of physical layer multicasting (i.e. \textit{PHY multicasting}) \cite{multicast}-\cite{multicast-jindal}. Since a single data stream
is sent to all receivers, there is no multiuser interference.
In the remainder of this paper, this case will be referred to as multicast
transmission.
The second service type is known as unicast, in which a transmitter
has an individual message for each receiver. Due to the nature of the wireless medium
and the use of multiple antennas, multiple simultaneous unicast transmissions are possible in the downlink of a base station (BS). In these cases, multiple streams are
simultaneously sent, which motivates precoding techniques that mitigate the
multiuser interference. In information theory terms, this
service type has been studied using the broadcast channel \cite{haardt}. In
the remainder of this paper, this case will be referred to as \textit{downlink} transmission.\smallskip

In the literature, the precoding techniques for downlink tranmissions can be further classified as\cite{maha_TSP}:
\begin{enumerate}
\item \textit{Group-level precoding} in which multiple codewords are transmitted simultaneously
but each codeword is addressed to a group of users. This case is also known as multigroup
multicast precoding \cite{g-multicast}-\cite{silva} and the precoder design
is dependant on the channels in each user group.
\item \textit{User-level precoding} in which multiple codewords are transmitted simultaneously
but each codeword is addressed to a single user. This case is also known as multiantenna
broadcast channel precoding \cite{mats}-\cite{ghaffar} and the precoder design depends on the channels of the individual users. This is a special
case of group level precoding where each group consists of a single user.
\item \textit{Symbol-level precoding} in which multiple symbols are transmitted simultaneously
and each symbol is addressed to a single user 
\cite{Christos-1}-\cite{maha_TSP}. This is also known as a constructive interference
precoding and the precoder design is dependent on  both the channels (CSI) and
the symbols of the users (DI).
\end{enumerate} 

The main idea  of symbol-based precoding is to jointly utilize the spatial cross-coupling between  the users' channel and the received symbols which depend on both channel state and transmitted symbols. When untreated, this cross-coupling leads to interference among the symbol streams of the users. Several spatial processing techniques decouple  the multiuser transmission to reduce the interference power received at each terminal\cite{haardt}. On the other hand, \cite{Christos-1} classifies the interference in the scenario of BPSK and QPSK into two types: constructive and destructive. Based on this classification, a selective channel inversion scheme is proposed to cancel the destructive interference while retaining the constructive one to be received at the users' terminal. A more elaborated scheme is proposed in \cite{Christos}, which  rotates the destructive interference to be received as useful signal with the constructive one. These schemes outperform the conventional precoding schemes \cite{haardt} and show considerable gains. However, the anticipated
gains come at the expense of additional complexity at the system design level. Assuming
that
the channel coherence time is $\tau_{c}$, and the symbol period is $\tau_s$, with $\tau_c\gg\tau_s$ for slow fading
channels, the user precoder has to be recalculated with a frequency of ${\tau_c}^{-1}$
in comparison with the symbol based precoder ${\min(\tau_c,\tau_s)}^{-1}={\tau_s}^{-1}$. Therefore, faster precoder calculation and switching is needed in
the symbol-level precoding which can be translated to more complex and expensive hardware.\smallskip

 In this direction, \cite{maha}-\cite{maha_TSP} have proposed a symbol based precoding to exploit the interference by establishing the connection between the constructive interference precoding and multicast. 
Moreover, several constructive interference precoding schemes have been proposed in \cite{maha_TSP}, including Maximum ratio transmission (MRT)-based algorithm  and objective-driven constructive interference techniques. The MRT based algorithm, titled as Constructive interference MRT (CIMRT), exploits the singular value decomposition (SVD) of the concatenated channel matrix. This enables the decoupled rotation using Givens rotation matrices  between the users' channels subspaces to ensure that the interference is received constructively at the users. On the other hand, the objective- driven optimization formulates the constructive interference problem by considering its relation to PHY-multicasting. Many metrics are addressed such as minimizing transmit power, maximizing the minimum SNR and maximizing the sum rate. However, the aforementioned precoding techniques design the transmitted signal so that it is received exactly the desired constellation point.

In the current paper, we aim at optimizing the constructive interference among the spatial streams while we allow for  more flexible precoding design.  We exploit the fact that the received symbol should lie in the correct detection region but not necessarily at the exact constellation point. This provides flexibility at the precoding design level in comparison with \cite{maha}-\cite{maha_TSP}, where the precoding is designed to target the exact constellation point (see Fig. \ref{qpsk}). As shown herein, this flexibility can be translated into more energy-efficient transmissions.   \smallskip

The contribution of the paper can be summarized as:
\begin{itemize}
\item Based on the constructive interference definition\cite{maha_TSP} and the relaxed detection region concept, we propose a symbol level precoding that minimizes the transmit power subject to SNR target constraints and maximizes the minimum SNR subject to total transmit power.  
\item The impact of relaxed detection on the symbol error rate and consequently the effective rate is analyzed. 
\item The tradeoff between SER increase (effective rate) and the transmit power saving is investigated by exploiting an energy efficiency metric.
\end{itemize}

The rest of the paper is organized as follows: the channel and  system
 model are explained in section (\ref{system}), while section (\ref{constructive}) revisits the definition of constructive interference. Section (\ref{powmin}) exploits the constructive interference with relaxed detection in symbol-level precoding that minimizes the transmit power subject to SNR target. 
 Moreover, the problem of maximizing the minimum SINR is tackled in section (\ref{maxmins}). The symbol error rate performance is studied in section (\ref{SERsection}). The impact of the increased error resulting from relaxed detection region on the effective rate is discussed by studing the energy efficiency metric in section (\ref{tradeoff}). Finally, the performance
of the proposed
algorithms is evaluated in section (\ref{sim}).\smallskip

\textbf{Notation}:  We use boldface upper and lower case letters for
 matrices and column vectors, respectively. $(\cdot)^H$, $(\cdot)^*$
 stand for Hermitian transpose and conjugate of $(\cdot)$. $\mathbb{E}(\cdot)$ and $\|\cdot\|$ denote the statistical expectation and the Euclidean norm,  $\mathbf{A}\succeq \mathbf{0}$ is used to indicate the positive
semidefinite matrix. $\angle(\cdot)$, $|\cdot|$ are the angle and magnitude  of $(\cdot)$ respectively. $\mathcal{R}(\cdot)$, $\mathcal{I}(\cdot)$
 are the real and the imaginary part of $(\cdot)$, $i$ indicates the complex part of the nuumber. Finally, the vector of
 all zeros with length of $K$ is defined as $\mathbf{0}^{K\times 1}$.
\vspace{-0.15cm} 
\section{System and Signal Models}
\label{system}
%\vspace{-0.1cm} 
We consider a single-cell multiple-antenna downlink scenario,
where a single BS is equipped with $M$
transmit antennas that serves $K$ user terminals,
each one of them equipped with a single receiving antenna. The adopted
modulation technique is M-PSK.
We assume a quasi static block fading channel $\mathbf{h}_j\in\mathbb{C}^{1\times
M}$ between
the BS antennas and the $j^{th}$ user, where the received signal at
j$^{th}$ user is
written as
%\vspace{-0.3cm}
\begin{eqnarray}
y_j[n]&=&\mathbf{h}_j\mathbf{x}[n]+z_j[n].
\end{eqnarray} $\mathbf{x}[n]\in\mathbb{C}^{M\times 1}$ is the transmitted symbol sampled signal vector at time $n$ from the multiple antennas
transmitter and  $z_j$ denotes the noise at $j^{th}$ receiver, which is assumed i.d.d  complex Gaussian distributed variable $\mathcal{CN}(0,1)$. A compact formulation
of the received signal at all users' receivers can be written as
\vspace{-0.1cm}
\begin{eqnarray}
\mathbf{y}[n]&=&\mathbf{H}\mathbf{x}[n]+\mathbf{z}[n].
\end{eqnarray}
Let $\mathbf{x}[n]$ be written as $\mathbf{x}[n]=\sum^K_{j=1}\sqrt{p_j[n]}\mathbf{w}_j[n]d_j[n]$,
where $\mathbf{w}_j$ is the $\mathbb{C}^{M\times
1}$ unit power precoding vector for the user $j$. The received signal at $j^{th}$
user ${y}_j$ in $n^{th}$ symbol period is given by
\begin{eqnarray}
\label{rx_o}
\hspace{-1cm}{y}_j[n]=\sqrt{p_j[n]}\mathbf{h}_j\mathbf{w}_j[n] d_j[n]+\displaystyle\sum_{k\neq j}\sqrt{p_k[n]}\mathbf{h}_j\mathbf{w}_k[n]
d_k[n]+z_j[n]
\end{eqnarray}
where $p_j$ is the allocated power to the $j^{th}$ user. A more detailed compact system formulation
is obtained by stacking the received signals and the noise
components for the set of K selected users as
\begin{eqnarray}
\mathbf{y}[n]=\mathbf{H}\mathbf{W}[n]\mathbf{P}^{\frac{1}{2}}[n]\mathbf{d}[n]+\mathbf{z}[n]
\end{eqnarray}
with $\mathbf{H} = [\mathbf{h}_1,..., \mathbf{h}_K]^T \in\mathbb{C}^{K\times M} $, $\mathbf{W}=[\mathbf{w}_1, ...,\mathbf{w}_K]\in\mathbb{C}^{nt\times M}$ as the
compact channel and precoding matrices. Notice that the transmitted signal $\mathbf{d}\in\mathbb{C}^{K\times 1}$
includes the uncorrelated data symbols $d_k$ for all users with $\mathbb{E}[{|d_k|^2}] = 1$, $\mathbf{P}^{\frac{1}{2}}[n]$
is the power allocation matrix $\mathbf{P}^{\frac{1}{2}}[n]=diag(\sqrt{p_1[n]},\hdots,\sqrt{p_K[n]})$.
It should be noted that both CSI and data information (DI) are available at the transmitter side. From now on, we assume that the precoding design is performed at each symbol period and accordingly we drop the time index for the sake of notation.
\subsection{Power constraint}
In the conventional user-level precoding, the transmitter needs to precode every $\tau_{c}$
which means that the power constraint has to be satisfied along the coherence time
$\mathbb{E}_{\tau_c}\{\|\mathbf{x}\|^2\}\leq
P$. Taking the expectation of $\mathbb{E}_{\tau_c}\{\|\mathbf{x}\|^2\}=\mathbb{E}_{\tau_c}\{tr(\mathbf{W}\mathbf{d}\mathbf{d}^H\mathbf{W}^H)\}$,
and since $\mathbf{W}$ is fixed along $\tau_c$, the previous expression can
be reformulated as $tr(\mathbf{W}\mathbb{E}_{\tau_c}\{\mathbf{d}\mathbf{d}^H\}\mathbf{W}^H)=tr(\mathbf{W}\mathbf{W}^H)=\sum^K_{j=1}\|\mathbf{w}_j\|^2$,
where $\mathbb{E}_{\tau_c}\{\mathbf{d}\mathbf{d}^H\}=\mathbf{I}$ due to uncorrelated
symbols over $\tau_c$.

However, in symbol level precoding the power constraint should be guaranteed
for each symbol vector transmission namely for each $\tau_s$. In this case
the power constraint equals to $\|\mathbf{x}\|^2=\mathbf{W}\mathbf{d}\mathbf{d}^H\mathbf{W}^H=\|\sum^K_{j=1}\mathbf{w}_jd_j\|^2$.
In the next sections, we characterize the constructive interference and show
how to exploit it in the multiuser downlink transmissions.

 \section{Constructive Interference}
 \label{constructive}
 \vspace{-0.1cm}
 
In general, the interference diverts the desired constellation point randomly in any direction, which possibly pushes it outside the correct detection region.
% The interference is a random deviation which can move the desired constellation point in any direction and possibly outside of the correct detection
% region. 
To address this problem, the power of the interference has been used in the past to regulate its effect on the desired signal point. 
 However,
in symbol level precoding (e.g. M-PSK) this interference can be constructed in advance in order to push the received symbols further into the correct detection region and, as a consequence it enhances the system performance. Therefore, the interference can
be classified into constructive or destructive based on whether it facilitates or deteriorates the correct detection of the received symbol. For BPSK and QPSK scenarios, a detailed classification of interference is discussed thoroughly in \cite{Christos-1}. The required conditions to have constructive interference for any M-PSK modulation have been described in \cite{maha_TSP}, but we mention here the definition of constructive interference for the sake of completeness.
\vspace{-0.1cm}
\subsection{Constructive Interference Definition}

Assuming both DI and CSI are available at the transmitter, the unit-power
 created interference from the $k^{th}$ data stream on $j^{th}$ user can be formulated as:
\vspace{-0.1cm}
\begin{equation}
\rho_{jk}=\frac{\mathbf{h}_{j}\mathbf{w}_k}{\|\mathbf{h}_{j}\|\|\mathbf{w}_k\|}.
\end{equation}
Since the adopted modulations are M-PSK ones, a definition for
constructive interference can be stated as:\smallskip

\begin{newtheorem}*{lemma}{\textbf{Lemma}\cite{maha_TSP}}
\begin{lemma}
\label{lemma}
An M-PSK modulated symbol $d_k$ is said to receive constructive
interference from another simultaneously transmitted symbol $d_j$ which is
associated with $\mathbf{w}_j$ if and only if the following inequalities hold   
\begin{equation}\nonumber
\label{one}
\angle{d_j}-\frac{\pi}{M}\leq \arctan\Bigg(\frac{\mathcal{I}\{\rho_{jk}d_{k}\}}{\mathcal{R}\{\rho_{jk}d_{k}\}}\Bigg)\leq \angle{d_j}+\frac{\pi}{M},
\end{equation}
\begin{equation}\nonumber
\label{two}
\mathcal{R}\{{d_k}\}.\mathcal{R}\{\rho_{jk}
d_{j}\}>0, \mathcal{I}\{{d_k}\}.\mathcal{I}\{\rho_{jk}d_{j}\}>0.\\
\end{equation}
\end{lemma}\smallskip
 \vspace{0.1cm}

\end{newtheorem}

\begin{newtheorem}*{cor}{\textbf{Corollary}\cite{maha_TSP}}
\begin{cor}
The constructive interference is mutual.
If the symbol $d_j$ constructively interferes with $d_k$, then
the interference from transmitting  the symbol $d_k$ 
is constructive to $d_j$.\\ 
\end{cor}

\end{newtheorem}

 \section{Constructive Interference for Power Minimization}
 \label{powmin}
 \subsection{Constructive Interference Power Minimization Precoding (CIPM) with Strict Constellation Targets \cite{maha_TSP}}

  From the definition of constructive interference, we should design the constructive interference precoders by granting that the sum of the
precoders and data symbols forces the received
signal to an exact MPSK constellation point namely an exact phase for each user. 
 Therefore, the optimization that
 minimizes the transmit power and grants
 the constructive reception of the transmitted data symbols can be written
 as 
\vspace{-0.2cm}
\begin{eqnarray}
\label{powccm}
\hspace{-0.4cm}\mathbf{w}_k(d_j,\mathbf{H},\boldsymbol\zeta)
\hspace{-0.1cm}&=&\arg\underset{\mathbf{w}_1,\hdots,\mathbf{w}_K}{\min}\quad \|\sum^K_{k=1}\mathbf{w}_kd_k\|^2\\\nonumber
\hspace{-0.1cm}&s.t.&\begin{cases}\mathcal{C}1:\angle(\mathbf{h}_j\sum^K_{k=1}\mathbf{w}_k
d_k)=\angle(d_j),
\forall j\in K\\
\mathcal{C}2:\|\mathbf{h}_j\sum^K_{k=1}\mathbf{w}_kd_k\|^2\geq\sigma^2\zeta_j\quad
, \forall j\in K,
\end{cases}
\end{eqnarray}\\
where $\zeta_j$ is the SNR target for the $j^{th}$ user, and ${\boldsymbol\zeta}=[\zeta_1,\hdots,\zeta_K]$ is the vector that contains all the SNR targets.  The set of constraints $\mathcal{C}_1$
gaurantees that the received signal for each user has the correct phase so that the right MPSK symbol $d_j$ can be detected. 

% The optimization
% can be rewritten as
% \begin{eqnarray}\nonumber
% \label{unitrank}
% \hspace{-0.1cm}\mathbf{x}(d_j,\mathbf{H},\boldsymbol\zeta)
% &=&\arg\underset{\mathbf{x}}{\min}\quad \|\mathbf{x}\|^2\\\nonumber
% &s.t.&\begin{cases}\mathcal{C}1:\angle(\mathbf{h}_j\mathbf{x})=\angle(d_j),
% \forall j\in K\\
% \mathcal{C}2:\|\mathbf{h}_j\mathbf{x}\|^2\geq\sigma^2\zeta_j\quad
% , \forall j\in K.
% \end{cases}
% \end{eqnarray}\\

\begin{figure*}[t]
\vspace{-0.5cm}
\hspace{0.2cm}
\begin{tabular}[t]{c}
\begin{minipage}{17 cm}
%\vspace{-1cm}
 \begin{center}
\hspace{-.5cm}\includegraphics[scale=0.3]{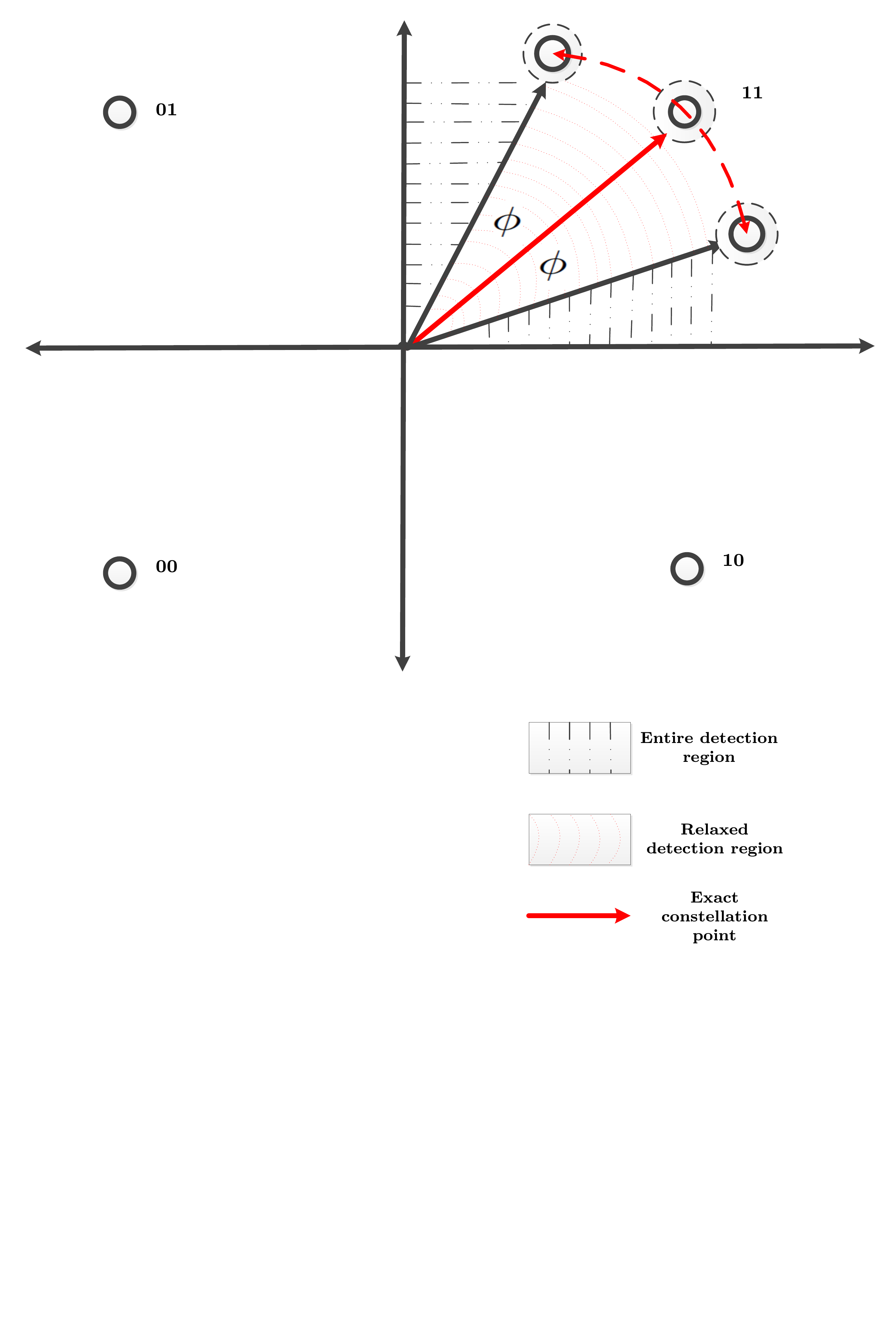}
\vspace{-4.2cm}\caption{\label{qpsk} Constructive Interference Symbol Level Precoding  in  Multiuser MISO Based on Relaxed Detection Region. The phase $\phi$ delimits the relaxed region. }
\end{center}
\end{minipage}\\
\hline
\end{tabular}
\end{figure*}

\begin{newtheorem}{thm1}{\textbf{Theorem}} 

\end{newtheorem}

The solution of (\ref{powccm}) is  fully derived in\cite{maha}-\cite{maha_TSP}, and can be written as
%\vspace{-0.2cm}
\hspace{-0.2cm}\begin{eqnarray}\nonumber
\label{CIPM}
\hspace{-0.5cm}\mathbf{x}&=&\sum^K_{k=1}\mathbf{w}_kd_k=-0.5i\sum^K_{j=1}\mu_j\mathbf{h}^H_j
-0.5\sum_j\alpha_j\mathbf{h}^H_j\\
 &\equiv&\sum^K_{j=1}\nu_j\mathbf{h}^H_j,\forall
i\in K
\end{eqnarray}
where $\nu_j\in \mathbb{C}=-0.5i\mu_j-0.5\alpha_j$. The optimal values of the Lagrangian variables $\mu_j$ and $\alpha_j$ can
be found by  solving the set of $2K$ equations (\ref{setoos}).
 The final constrained 
constellation
precoder can be found by substituting all $\mu_j$ and $\alpha_j$ in (\ref{CIPM}). 
\begin{figure*}[t]
\vspace{-0.2cm}
\hspace{0.2cm}
\begin{tabular}[t]{c}
\begin{minipage}{17 cm}
%\vspace{-1cm}
 \begin{eqnarray}
\label{multicasteq}
\begin{array}{cccc}
\label{setoos}
0.5K\|\mathbf{h}_1\|(\sum_k(-\mu_k+\alpha_ki)\|\mathbf{h}_k\|\rho_{1k}&-&\sum_k(-\mu_k+\alpha_ki)\|\mathbf{h}_k\|\rho^{*}_{1k})=\sqrt{\zeta^{}_{1}}\mathcal{I}(d_1)\\
0.5K\|\mathbf{h}_1\|(\sum_k(-\mu_ki-\alpha_k)\|\mathbf{h}_k\|\rho_{1k}&+&\sum_k(-\mu_ki-\alpha_k)\|\mathbf{h}_k\|\rho^{*}_{1k})=\sqrt{\zeta^{}_{1}}\mathcal{R}(d_1)\\
\quad&\vdots&\\
0.5K\|\mathbf{h}_K\|(\sum_k(-\mu_k+\alpha_ki)\|\mathbf{h}_k\|\rho_{Kk}&-&\sum_k(-\mu_k+\alpha_ki)\|\mathbf{h}_k\|\rho^{*}_{Kk})=\sqrt{\zeta_{K}}\mathcal{I}(d_K)\\
0.5K\|\mathbf{h}_K\|(\sum_k(-\mu_ki-\alpha_k)\|\mathbf{h}_k\|\rho_{Kk}&+&\sum_k(-\mu_ki-\alpha_k)\|\mathbf{h}_k\|\rho^{*}_{Kk})=\sqrt{\zeta_{K}}\mathcal{R}(d_K)\\
\end{array}
\end{eqnarray}
\end{minipage}\\
%\vspace{-0.3cm}\\
\hline
\hline
\end{tabular}
\end{figure*}

\begin{cor}
The constructive interference precoding must span the subspaces of all involved users.
\end{cor}

\subsection{Proposed Constructive Interference Power Minimization precoding with Relaxed Detection Region}
To grant a correct M-PSK symbol detection, the  received symbol should lie in the correct
detection region. Fig. (\ref{qpsk}) depicts the detection region of the QPSK symbol $\frac{1+i}{\sqrt{2}}$ which spans the phases $[0^{\circ},90^{\circ}]$.
In the previous section, we designed the transmitted symbol to be received with the exact phase
of the target data
symbols except the random deviation resulting from the noise at the receiver. On the other hand, the same symbol can be correctly detected as $``11"$  within a range of phases as long as it lies in the first quadrant and the receiver noise does not push it outside the detection region. Therefore,
it is not necessary to design precoding so that it aims the exact phase, but the targeted receive symbol can span the range of $[45-\phi, 45+\phi]$ with $0\leq\phi\leq 45$, where $\phi$ denotes the phase margin of the relaxed detection region.
Therefore, more flexibility for the system design can be obtained and higher gains can be harnessed as shown in the numerical results section. It should be noted that the above example is for QPSK, but the concept is straightforwardly extendable to other MPSK modulations.
Since the detection region of symbols span different phases, we can utilize
this property by relaxing the transmitted constellation point to include this angular span, which is called the \textit{relaxed detection region}. The relaxed optimization
can be formulated as
\vspace{-0.4cm}
\hspace{-0.4cm}\begin{eqnarray}\nonumber
\label{CIPMR}
&\hspace{-0.2cm}\mathbf{w}_{j}(\mathbf{H},\mathbf{d},\boldsymbol{\zeta},\mathbf{\Phi}_1,\mathbf{\Phi}_2)&\hspace{-0.1cm}=\arg\underset{\mathbf{w}_{j}}{\min}\quad\|\sum^K_{j=1}\mathbf{w}_{j}d_j\|^2\\\nonumber
&\hspace{-4.0cm}s.t.&\hspace{-3.0cm}\begin{cases}\mathcal{C}_1:\angle(\underset{\phi^{'}_{j1}}{\underbrace{d_j-\phi_{j1}}})\leq\angle(\mathbf{h}_j\sum^K_{j=1}\mathbf{w}_{j})\leq\angle(\underset{\phi^{'}_{j2}}{\underbrace{d_j+\phi_{j2}}}),\forall j\in K\\
\mathcal{C}_2:\|\mathbf{h}_j\sum^K_{j=1}\mathbf{w}_{j}\|^2\geq\sigma^2\zeta_j\quad
\forall j\in K.\end{cases}\\
\end{eqnarray}
If we use $\mathbf{x}=\sum_{j=1}\mathbf{w}_jd_j$, the problem can be expressed as
\begin{eqnarray}\nonumber
\label{powminr}
&\mathbf{x}&(\mathbf{H},\mathbf{d},\boldsymbol\zeta,\mathbf{\Phi}_1,\mathbf{\Phi}_2)\hspace{-0.1cm}=\arg\underset{\mathbf{x}}{\min}\quad\|\mathbf{x}\|^2\\\nonumber
&s.t.&\begin{cases}\mathcal{C}_1:\angle(\underset{\phi^{'}_{j1}}{\underbrace{d_j-\phi_{j1}}})\leq\angle(\mathbf{h}_j\mathbf{x})\leq\angle(\underset{\phi^{'}_{j2}}{\underbrace{d_j+\phi_{j2}}}),\forall j\in K\\
\mathcal{C}_2:\|\mathbf{h}_j\mathbf{x}\|^2\geq\sigma^2\zeta_j\quad
\forall j\in K.\end{cases}
\end{eqnarray}
where $\phi_{j1}$ and $\phi_{j2}$ are the phase thresholds that received symbols
should lie in without the noise drifting, $\boldsymbol{\phi}_1$ and $\boldsymbol{\phi}_2$
are the vectors that 
contain all $\phi_{j1}$ and $\phi_{j2}$ respectively. Although this relaxes the phase constraints on the constructive interference design,
it increases the system susceptibility to noise. Therefore, this phase margin should
be related to the SNR targets to guarantee certain power saving and SER by  selecting  the allowable values of $\phi_{j1}$ and $\phi_{j2}$. The optimization can be written\footnote{$\pm$ in ($\mathcal{C}_2$-\ref{CIPMR}) indicates that the sign can be positive or negative depending on the value of $\sin\phi$  function}
\begin{eqnarray}
\label{relaxedp}
&\mathbf{x}&(\mathbf{H},\mathbf{d},\boldsymbol{\zeta},\mathbf{\Phi}_1,\mathbf{\Phi}_2)=\arg\underset{\mathbf{x}}{\min}\quad\|\mathbf{x}\|^2\\\nonumber
&s.t.&\begin{cases}
\mathcal{C}_1:\mathbf{h}_j\mathbf{x}+\mathbf{x}^H\mathbf{h}^H_j\gtreqless2\sqrt{\zeta_j}u_j,\forall j\in K \\
\mathcal{C}_2:\mathbf{h}_j\mathbf{x}-\mathbf{x}^H\mathbf{h}^H_j\gtreqless\pm2i\sqrt{\zeta_j}\sqrt{1-u^2_j},\forall j\in K\\
\mathcal{C}_3:\cos(\phi^{'}_{j2})\leq u_j\leq \cos(\phi^{'}_{j1}), \forall j \in K
\end{cases}
\end{eqnarray}
where $u_j$ is an auxiliary variable. This optimization has $3K$ constraints that need to be satisfied. The Lagrangian for this problem can be written as 
\begin{eqnarray}\nonumber\label{q}\mathcal{L}(\mathbf{x})&=&\|\mathbf{x}_r\|^2+\sum_j\alpha_j(\mathbf{h}_j\mathbf{x}+\mathbf{x}^H\mathbf{h}^H_j-2\sqrt{\zeta_j}u_j)\\\nonumber
&+&\sum_j\mu_j(\mathbf{h}_j\mathbf{x}-\mathbf{x}^H\mathbf{h}^H_j-2i\sqrt{\zeta_j}\sqrt{1-u^2_j})\\\nonumber
&+&\sum_{j}\alpha_j(u_j-\cos(\phi^{'}_{j,1})))+\sum_{j}\gamma_j(u_j-\cos(\phi^{'}_{j,2}))).
\end{eqnarray}

\begin{figure*}[t]
%\vspace{-0.2cm}
\hspace{0.2cm}
\begin{tabular}[t]{c}
% \begin{minipage}{17 cm}
% \vspace{-0.5cm}
% % \begin{eqnarray}\nonumber
% % \label{R_detection}
% % \hspace{-0.7cm}\mathcal{L}(\mathbf{x})&=&\|\mathbf{x}\|^2+\sum^K_j\alpha_j\Bigg({\bigg(\mathbf{h}_j\mathbf{x}-\mathbf{x}^H\mathbf{h}^H}-i\big(\mathbf{h}_j\mathbf{x}+(\mathbf{h}_j\mathbf{x})^H\big)\bigg)\tan(\theta_1)\Bigg)\\\nonumber
% % &+&\sum^K_j\mu_j\bigg(\mathbf{h}_j\mathbf{x}-(\mathbf{h}_j\mathbf{x})^H-i(\mathbf{h}_j\mathbf{x}+(\mathbf{h}_j\mathbf{x})^H)\tan(\theta_2)\bigg)+-0.5i\sum^K_j\eta_j\bigg(\mathbf{h}_j\mathbf{x}\\\nonumber
% % &-&(\mathbf{h}_j\mathbf{x})^H-\mathcal{I}\{\sqrt{\sigma^2\zeta_j}(\angle(d_j+\phi_{j1})\}\bigg)+0.5\sum^K_j\varrho_j\bigg(\mathbf{h}_j\mathbf{x}+(\mathbf{h}_j\mathbf{x})^H-\mathcal{R}\{\sqrt{\sigma^2\zeta_j}(\angle(d_j+\phi_{j1})\}\bigg)\\
% % \end{eqnarray}

% \end{minipage}\\
%\vspace{-0.3cm}\\
\hline
\hline
\end{tabular}
\end{figure*}
Differentiating
$\mathcal{L}(\mathbf{x})$ with respect to $\mathbf{x}^*$ and $u_j$ yields:
\begin{eqnarray}\nonumber
\vspace{-0.5cm}\frac{d\mathcal{L}(\mathbf{x},u_j)}{d\mathbf{x}}&=&\mathbf{x}+\sum_{j}\alpha_j\mathbf{h}^H_j-\sum_{j}\mu_j\mathbf{h}^H_j,\\
\frac{d\mathcal{L}(\mathbf{x},u_j)}{du_j}&=&-2\sqrt{\zeta_j}+2i\sqrt{\zeta_j}\frac{u_i}{\sqrt{1-u^2_i}}+\lambda_j+\gamma_j.
\end{eqnarray}
By equating $\frac{d\mathcal{L}(\mathbf{x},u_i)}{d\mathbf{x}^*}=0$ and $\frac{d\mathcal{L}(\mathbf{x},u_i)}{d\mathbf{u}_i}=0$, we can get the following expressions
\begin{eqnarray}
\label{rw1}
\mathbf{x}=\sum_j-\alpha_j\mathbf{h}^H_j+\mu_j\mathbf{h}^H_j
\end{eqnarray}
\begin{eqnarray}
\label{rw2}
u_j=\pm\frac{2\sqrt{\zeta_j}-\lambda_j-\gamma_j}{\sqrt{-4\sqrt{\zeta_j}(\lambda_j+\gamma_j)+\lambda^2_j+2\lambda_j\gamma_j+\gamma^2_j}}.
\end{eqnarray}
Substituting (\ref{rw1})-(\ref{rw2}) in the constraints, we have the set of inequalities
(\ref{multicasteq}). It can be noted that the solution of (\ref{powccm}) is a special case of (\ref{relaxedp}) when $\phi_{j1}$ and $\phi_{j2}$ are equal
to zero.\smallskip

\begin{figure*}[t]
\begin{tabular}[t]{c}
\begin{minipage}{19 cm}
 \begin{eqnarray}
\label{multicasteq}
\begin{array}{cccc}
\label{setoo}
\hspace{-2.2cm}0.5\|\mathbf{h}_1\|(\sum_k(-\mu_k+\alpha_ki)\|\mathbf{h}_k\|\rho_{1k}&-&\sum_k(-\mu_k+\alpha_ki)\|\mathbf{h}_k\|\rho^{*}_{1k})=\sqrt{\zeta^{}_{1}}\sqrt{1-\frac{(2\sqrt{\zeta_1}-\lambda_1-\gamma_1)^2}{{-4\sqrt{\zeta_1}(\lambda_1+\gamma_1)+\lambda^2_1+2\lambda_1\gamma_1+\gamma^2_1}}}\\
\hspace{-2.2cm}0.5\|\mathbf{h}_1\|(\sum_k(-\mu_ki-\alpha_k)\|\mathbf{h}_k\|\rho_{1k}&+&\sum_k(-\mu_ki-\alpha_k)\|\mathbf{h}_k\|\rho^{*}_{1k})=\sqrt{\zeta^{}_{1}}\frac{2\sqrt{\zeta_1}-\lambda_1-\gamma_1}{\sqrt{-4\sqrt{\zeta_1}(\lambda_1+\gamma_1)+\lambda^2_1+2\lambda_1\gamma_1+\gamma^2_1}}\\
\quad&\vdots&\\
\hspace{-2.2cm}0.5\|\mathbf{h}_K\|(\sum_k(-\mu_k+\alpha_ki)\|\mathbf{h}_k\|\rho_{Kk}&-&\sum_k(-\mu_k+\alpha_ki)\|\mathbf{h}_k\|\rho^{*}_{Kk})=\sqrt{\zeta_{K}}\sqrt{1-\frac{(2\sqrt{\zeta_K}-\lambda_K-\gamma_K)^2}{{-4\sqrt{\zeta_K}(\lambda_K+\gamma_K)+\lambda^2_K+2\lambda_K\gamma_K+\gamma^2_K}}}\\
\hspace{-2cm}0.5\|\mathbf{h}_K\|(\sum_k(-\mu_ki-\alpha_k)\|\mathbf{h}_k\|\rho_{Kk}&+&\sum_k(-\mu_ki-\alpha_k)\|\mathbf{h}_k\|\rho^{*}_{Kk})=\sqrt{\zeta_{K}}\frac{2\sqrt{\zeta_K}-\lambda_K-\gamma_K}{\sqrt{-4\sqrt{\zeta_K}(\lambda_K+\gamma_K)+\lambda^2_K+2\lambda_K\gamma_K+\gamma^2_k}}\\
\hspace{-2cm}{2\sqrt{\zeta_1}-\lambda_1-\gamma_1}&\leq& {\sqrt{-4\sqrt{\zeta_1}(\lambda_1+\gamma_1)+\lambda^2_1+2\lambda_1\gamma_1+\gamma^2_1}} \cos(\phi^{'}_{11})\\
\hspace{-2.2cm}{2\sqrt{\zeta_1}-\lambda_1-\gamma_1}&\geq& {\sqrt{-4\sqrt{\zeta_1}(\lambda_1+\gamma_1)+\lambda^2_1+2\lambda_1\gamma_1+\gamma^2_1}}\cos(\phi^{'}_{12})\\
\quad&\vdots&\\
\hspace{-2.2cm}{2\sqrt{\zeta_K}-\lambda_K-\gamma_K}&\leq& {\sqrt{-4\sqrt{\zeta_K}(\lambda_K+\gamma_K)+\lambda^2_K+2\lambda_K\gamma_K+\gamma^2_K}}\cos(\phi^{'}_{K1})\\
\hspace{-2.2cm}{2\sqrt{\zeta_K}-\lambda_K-\gamma_K}&\geq& {\sqrt{-4\sqrt{\zeta_K}(\lambda_K+\gamma_K)+\lambda^2_K+2\lambda_K\gamma_K+\gamma^2_K}}\cos(\phi^{'}_{K2})
\end{array}
\end{eqnarray}
\normalsize
\end{minipage}\\
\vspace{-0.3cm}\\
\hline
\hline
\end{tabular}
\end{figure*}
\subsubsection{Equal phase margin solution for power minimization}
A simpler solution can be found for the scenario of $\phi_{j1}=\phi_1,\forall j\in K$ and $\phi_{j2}=\phi_2,\forall j\in K$ and $\phi=\phi_1=\phi_2$ by searching all the phases that lie in the relaxed region. The linear search is performed on the value of $\phi_u$ which is varied from $\angle d_j-\phi_1$ to $\angle d_j+\phi_2$ to achieve the minimum power consumption. For each value $\phi_u\in [\angle d_j-\phi,\angle d_j+\phi]$, we solve the following optimization
\hspace{-0.3cm}\begin{eqnarray}
\label{relaxedp}
&\hspace{-0.2cm}\mathbf{x}&(\mathbf{H},\mathbf{d},\boldsymbol{\zeta},\phi_u)=\arg\underset{\mathbf{x}}{\min}\quad\|\mathbf{x}\|^2\\\nonumber
&s.t.&\begin{cases}
\mathcal{C}_1:\mathbf{h}_j\mathbf{x}+\mathbf{x}^H\mathbf{h}^H_j\gtreqless2\sqrt{\zeta_j}\cos(\phi_u),\forall j\in K \\
\mathcal{C}_2:\mathbf{h}_j\mathbf{x}-\mathbf{x}^H\mathbf{h}^H_j\gtreqless2i\sqrt{\zeta_j}\sin(\phi_u),\forall j\in K,\\
\end{cases}
\end{eqnarray}
To find the phase within the phase margin that has the minimum power consumption
\begin{eqnarray}
\phi^*={\arg}\underset{\phi_u}{\min}\|\mathbf{x}(\mathbf
{H},\mathbf{d},\boldsymbol\zeta,\phi_u)\|^2.
\end{eqnarray}

The relaxed detection region allows for a larger search space to find the optimal CI precoding that requires 
minimal power to achieve the target SNR. On the other hand, this transmit power reduction comes at the expense of increasing the probability of
 symbol error rate (SER) due to the expected noise deviation of the received symbols from their exact constellation which is analytically studied in section (\ref{SERsection}) and numerically section (\ref{sim}).

\begin{newtheorem}{lem}{\textbf{Lemma}}
\end{newtheorem}
% \begin{lem}
% The relaxed criteria
% \end{lem}
\vspace{-0.5cm}
\subsection{Constructive Interference Power Minimization Bounds}
In order to assess the performance of the proposed algorithm, we use
two theoretical upper bound as follows\cite{maha_TSP}:

\subsubsection{Genie-aided upper-bound}
This bound occurs when all multiuser transmissions are constructively
interfering by nature and without the need to optimize the output vector. The minimum transmit power for a system that exploits the constructive interference
on symbol basis can be found by the following theoretical bound. If we assume $\mathbf{W}=\mathbf{H}^{'}$, where $\mathbf{H}^{'}=[\frac{\mathbf{h}^H_1}{\|\mathbf{h}_1\|}, \hdots,\frac{\mathbf{h}^H_K}{\|\mathbf{h}_K\|}]$. By exploiting singular value decomposition (SVD)of $\mathbf{H}$. $\mathbf{V}^{'}$ is the power scaled of $\mathbf{V}$
to normalize each column in $\mathbf{W}$ to unity. The received signal can be as
\begin{eqnarray}
\label{svd}
\vspace{-0.2cm}
\mathbf{y}&=&\mathbf{H}\mathbf{W}\mathbf{d}={{\mathbf{S}\mathbf{V}\mathbf{D}\mathbf{D}^H\mathbf{V}^{'}}}{{\mathbf{S}^H}}\mathbf{P}^{1/2}\mathbf{d}.
\end{eqnarray} 
If we denote $\mathbf{G}=\mathbf{SVD}\mathbf{D}^H$ and $\mathbf{B}=\mathbf{S}^H$. Utilizing the reformulation of $\mathbf{y}$ in
(\ref{svd}), the received signal can be written as 
\vspace{-0.05cm}
\begin{eqnarray}
\label{rot}
y_j=\|\mathbf{g}_j\|\sum^K_{k=1}\sqrt{p_k}\xi_{jk}d_k,
\end{eqnarray}
%\vspace{-0.1cm}
where $\mathbf{g}_j$ is the $j^{th}$ row of the matrix $\mathbf{G}$, $\xi_{jk}=\frac{\mathbf{g}_j\mathbf{b}_k}{\|\mathbf{g}_j\|}$. 
 
\begin{thm1}
The genie-aided minimum transmit power in the downlink of multiuser MISO   system can be found by solving the following optimization
\begin{eqnarray}\nonumber
\label{pr}
\hspace{-0.1cm}{P}_{min}&=&\arg\underset{p_1,\hdots,p_K}{\min}\quad\sum^K_{k=1}p_k\\\nonumber
&s.t.&\|\mathbf{g}_k\|^2(|\xi_{kk}|^2{p_k}+\sum^K_{j=1,j\neq k}{p_j}|\xi_{kj}|^2)\geq{\zeta_k},\forall
k\in K.\\
\end{eqnarray}
%\begin{proof}

%\end{proof}
\end{thm1}

% \begin{proof}According to (\ref{rot}), the bound in (\ref{pr}) can be found if all users face a constructive interference
% with respect to the multiuser transmissions of all other streams $\angle(\xi_{jk}d_j)=\angle
% d_k,\forall k,\forall j$.
% \end{proof}
% This bound can be mathematically found by solving the problem (\ref{pr}) using linear programming
% techniques\cite{boyd}. \\

\subsubsection{Optimal Multicast} 
Based on theorem (2), a theoretical upperbound can be characterized. This bound occurs if we drop the phase alignment
constraint $\mathcal{C}_1$. The intuition of
using this technique is the complete correlation among the information that needs to be communicated (i.e. same symbol for all users). The optimal input covariance $\mathbf{Q}$ for power minimization in a multicast system can
be found as a solution of the following optimization
\vspace{-0.05cm} 
\begin{eqnarray}
\label{powm1}
&\underset{\mathbf{Q}:\mathbf{Q}\succeq 0}{\min}&\quad tr(\mathbf{Q})\quad s.t.\quad\mathbf{h}_j\mathbf{Q}\mathbf{h}^H_j\geq\zeta_j\quad,\forall j\in K.
\end{eqnarray}
This problem is thoroughly solved in \cite{multicast}. A tighter upperbound
can be found by imposing a unit rank constraint on $\mathbf{Q}$\cite{multicast-jindal},
to allow the comparison with the unit rank transmit power minimization constructive interference
precoding 
\begin{eqnarray}
\label{unitrankm}
\underset{\mathbf{Q}:\mathbf{Q}\succeq 0, \text{Rank}(\mathbf{Q})=1}{\min} tr(\mathbf{Q})\quad s.t.\quad\mathbf{h}_j\mathbf{Q}\mathbf{h}^H_j\geq\zeta_j\quad,\forall j\in K
\end{eqnarray}

\section{Weighted Max Min SINR Algorithm for Constructive Interference Precoding (CIMM) Based on Relaxed Detection}\label{maxmins}
The weighted max-min SINR precoding enhances the relative fairness in the system by maximizing
the worst user SINR. This problem has been discussed in various scenarios
 such as downlink transmissions\cite{shitz}, and multicast \cite{multicast}. The authors of \cite{shitz} have solved the problem using the bisection technique.  On the other hand,
 the authors in \cite{multicast} have tackled this problem by finding the relation
 between the min-power problem and max-min problem and formulating both problem
 as convex optimization ones.
 \cite{maha_TSP} utilize the constructive interference to enhance the fairness in terms of weighted SNR. The challenging aspects are the additional constraints which guarantee that the data have been detected correctly at
 the receivers. The constructive interference max-min problem can be formulated as
 
\begin{eqnarray}
\label{CIMM}
\hspace{-0.7cm}\mathbf{w}_k=&\underset{\mathbf{w}_k}{\max}\underset{j}{\min}&\Big\{\frac{\|\mathbf{h}_j\sum^K_{k=1}\mathbf{w}_kd_k\|^2}{r_j}\Big\}^K_{i=1}\\\nonumber
&\hspace{-0.3cm}{s.t.}&\begin{cases}\mathcal{C}1:\|\sum^K_{k=1}\mathbf{w}_kd_k\|^2\leq P\\\nonumber
\mathcal{C}2:\angle(\mathbf{h}_j\sum^K_{k=1}\mathbf{w}_kd_k)=\angle(d_j),\quad\forall j\in K.
\end{cases}
\end{eqnarray}
% \begin{eqnarray}
% \label{maxmin}
% \hspace{-1.7cm}\mathbf{w}_k=&\underset{\mathbf{w}_k}{\max}\underset{j}{\min}&\Big\{\frac{\|\mathbf{h}_j\sum^K_{k=1}\mathbf{w}_kd_k\|^2}{r_j}\Big\}^K_{i=1}\\\nonumber
% &\hspace{-0.3cm}{s.t.}&\begin{cases}\mathcal{C}1:\|\sum^K_{k=1}\mathbf{w}_kd_k\|^2\leq P\\\nonumber
% \mathcal{C}2:\angle d_j-\theta_j\leq\angle(\mathbf{h}_j\sum^K_{k=1}\mathbf{w}_kd_k)\leq\angle(d_j+\theta_j),\quad\forall j\in K.
% \end{cases}\\
% \end{eqnarray}
\begin{figure*}[t]
\begin{tabular}[t]{c}
\begin{minipage}{18 cm}
 \begin{eqnarray}
\label{maxminr}
\hspace{-0.7cm}\mathbf{w}_k=&\underset{\mathbf{w}_k}{\max}\underset{j}{\min}&\Big\{\frac{\|\mathbf{h}_j\sum^K_{k=1}\mathbf{w}_kd_k\|^2}{r_j}\Big\}^K_{i=1}\\\nonumber
&\hspace{-0.3cm}{s.t.}&\begin{cases}\mathcal{C}1:\|\sum^K_{k=1}\mathbf{w}_kd_k\|^2\leq P\\\nonumber
\mathcal{C}2:\angle d_j-\phi_{j1}\leq\angle(\mathbf{h}_j\sum^K_{k=1}\mathbf{w}_kd_k)\leq\angle(d_j+\phi_{j2}),\quad\forall j\in K.
\end{cases}
\end{eqnarray}
\normalsize
\begin{eqnarray}
\label{maxminq}
\hspace{-1.2cm}\mathbf{x}(\mathbf{H},\mathbf{d},\boldsymbol \zeta,\mathbf{\Phi}_1,\mathbf{\Phi}_2,\mathbf{r})=&\underset{\mathbf{x}}{\max}\underset{j}{\min}&\Big\{\frac{\|\mathbf{h}_j\mathbf{x}\|^2}{r_j}\Big\}^K_{i=1}\\\nonumber
&{s.t.}&\begin{cases}\mathcal{C}1:\|\mathbf{x}\|^2\leq P\\\nonumber
\mathcal{C}2:\angle(d_j-\phi_{j})\leq\angle(\mathbf{h}_j\mathbf{x})\leq\angle(d_j+\phi_{j2}),\quad\forall j\in K\\
\end{cases}\end{eqnarray}
\begin{eqnarray}
\label{maxmint}
&\underset{t,\mathbf{x}}{\max}&\quad t\\\nonumber
&s.t.&\begin{cases}\mathcal{C}1:\|\mathbf{x}\|^2\leq P\\
\mathcal{C}2:\tan(\angle
d_j-\phi_{j1})\leq\frac{\mathbf{h}_j\mathbf{x}-(\mathbf{h}_j\mathbf{x})^H}{i({\mathbf{h}_j\mathbf{x}+(\mathbf{h}_j\mathbf{x})^H})}\leq\tan(\angle
d_j+\phi_{j2}),
\forall
j\in K\\
\mathcal{C}3:\mathcal{R}\{d_j\}.\mathcal{R}\{\mathbf{h}_j\mathbf{x}\}\geq 0, \forall
j\in K\\
\mathcal{C}4:\mathcal{I}\{d_j\}.\mathcal{I}\{\mathbf{h}_j\mathbf{x}\}\geq 0, \forall
j\in K\\
\mathcal{C}5:\|\mathbf{h}_j\mathbf{x}\|^2\geq R_jt,\forall j\in K.
\end{cases}
%&\quad&\frac{\mathbf{h}_j\mathbf{\mathbf{w}}+(\mathbf{h}_j\mathbf{\mathbf{w}})^H}{2}=\sqrt{\zeta_j}\mathcal{R}\{d\},\forall
%j\in K.
\end{eqnarray}
\end{minipage}\\
\vspace{-0.3cm}\\
\hline
\hline
\end{tabular}
\end{figure*}
where $r_j$ denotes the requested SNR target for the $j^{th}$ user and $P$ is the total power that should be allocated to the users. If we
denote $\mathbf{x}=\sum^K_{j=1}\mathbf{w}_jd_j$, the previous optimization
can be expressed as (\ref{CIMM}).
% \begin{eqnarray}
% \label{maxminq}
% \hspace{-1.2cm}\mathbf{q}(\mathbf{d},\mathbf{H},\mathbf{r})=&\underset{\mathbf{q}}{\max}\underset{j}{\min}&\Big\{\frac{\|\mathbf{h}_j\mathbf{q}\|^2}{r_j}\Big\}^K_{i=1}\\\nonumber
% &{s.t.}&\begin{cases}\mathcal{C}1:\|\mathbf{q}\|^2\leq P\\\nonumber
% \mathcal{C}2:\angle(d_j-\theta_j)\leq\angle(\mathbf{h}_j\mathbf{q})\leq\angle(d_j+\theta_j),\quad\forall j\in K\\
% \end{cases}\end{eqnarray}
where $\mathbf{r}$ is the vector that contains all the weights $r_j$. In \cite{multicast}\cite{maha_TSP}, it has been shown
that the optimal output vector is a scaled version of the min-power solution. The weighted max-min SINR problem has been solved using bisection method over  $t\in[0,1]$. The max-min SNR problem with relaxed detection region can be formulated as (\ref{maxminr}). To solve the problem, (\ref{maxminr}) is rewritten as (\ref{maxminq})-(\ref{maxmint}).
%  \hspace{-1.9cm}\begin{eqnarray}
% \hspace{-1.9cm}&\underset{t,\mathbf{q}}{\max}&\quad t\\\nonumber
% &s.t.&\begin{cases}\mathcal{C}1:\|\mathbf{q}\|^2\leq P\\
% \mathcal{C}2:\tan(\angle
% d_j-\phi_{j1})\leq\frac{\mathbf{h}_j\mathbf{q}-(\mathbf{h}_j\mathbf{q})^H}{i({\mathbf{h}_j\mathbf{q}+(\mathbf{h}_j\mathbf{q})^H})}\leq\tan(\angle
% d_j+\phi_{j2}),
% \forall
% j\in K\\
% \mathcal{C}3:\mathcal{R}\{d_j\}.\mathcal{R}\{\mathbf{h}_j\mathbf{q}\}\geq 0, \forall
% j\in K\\
% \mathcal{C}4:\mathcal{I}\{d_j\}.\mathcal{I}\{\mathbf{h}_j\mathbf{q}\}\geq 0, \forall
% j\in K\\
% \mathcal{C}5:\|\mathbf{h}_j\mathbf{q}\|^2\geq R_jt,\forall j\in K.
% \end{cases}
% %&\quad&\frac{\mathbf{h}_j\mathbf{\mathbf{w}}+(\mathbf{h}_j\mathbf{\mathbf{w}})^H}{2}=\sqrt{\zeta_j}\mathcal{R}\{d\},\forall
% %j\in K.
% \end{eqnarray}

%\subsection{Max-min SINR and min-power relation}
%\vspace{-0.2cm}
\normalsize
\subsection{Equal phase margin solution for max-min SNR}
For the scenario of $\phi_{j1}=\phi_1,\forall j\in K$, $\phi_{j2}=\phi_2,\forall j\in K$ and $\phi=\phi_1=\phi_2$, a simple solution can be found by searching all the phases that lie in the relaxed region. A linear search procedure is performed on the value of $\phi$ which is varied from $\angle d_j-\phi_1$ to $\angle d_j+\phi_2$ to achieve the objective function. For each value $\phi_u\in [\angle d_j-\phi_1,\angle d_j+\phi_2]$, we solve the following optimization

\vspace{-0.5cm}
\begin{eqnarray}
\label{relaxedp}
t(\phi_u)&=&\arg\underset{t,\phi_u}{\max}\quad t\\\nonumber
&s.t.&\begin{cases}
\mathcal{C}_1:\mathbf{h}_j\mathbf{x}+\mathbf{x}^H\mathbf{h}^H_j\gtreqless2 r_j{t}\cos(\phi_u),\forall j\in K \\
\mathcal{C}_2:\mathbf{h}_j\mathbf{x}-\mathbf{x}^H\mathbf{h}^H_j\gtreqless2r_j{t}i\sin(\phi_u),\forall j\in K,\\
\mathcal{C}_3:\|\mathbf{x}\|^2\leq P
\end{cases}
\end{eqnarray}
to find the phase that achieve the objective function
\begin{eqnarray}
\phi^*={\arg}\underset{\phi_u}{\max}\quad t(\phi_u),
\end{eqnarray}
where $t(\phi_u)$ is a function that maps the max-min value with its respective phase.

\footnotesize
\begin{center}
\hspace{-0.5cm}\begin{tabular}{p{8.3cm}}
\hline
\textbf{A1}:$t,\phi$ search for max-min SINR for CI precoding (CIMMR)\\
\hline
\begin{itemize}
\item Search over $\phi_u\in[\angle d-\phi,\angle d +\phi]$. For each $\phi_u$, find
\begin{enumerate}
\item $m_1\rightarrow 0$
\hspace{0.5cm} $m_2\rightarrow 1$
\item Repeat
\item set $t_m=\frac{m_1+m_2}{2}$
\item solve (\ref{relaxedp}) with dropping $\mathcal{C}_3$ and substituting $t_m$ in place of $t$, set $\hat{P}=\|\mathbf{x}\|^2$
\item if $\hat{P}\leq P$\newline
 \hspace{1.0cm}then\quad $t_m \rightarrow t_1$\newline
\hspace{1.0cm}else\quad$t_m \rightarrow t_2$\newline
 Until $|\hat{P}-P|\leq \delta$\newline
\item $\phi^*=\arg\max t(\phi_u)$
\end{enumerate}
\item\hspace{0.2cm}Return $\phi^*$, $t_m(\phi^*)$
\end{itemize}\\
\hline
\end{tabular}
\end{center}
\normalsize

\subsubsection{Complexity of CIPMR and CIMMR}
The complexity of CIMM and CIPM are discussed in \cite{maha_TSP}. CIPMR and  CIMMR have additional complexity of $\log_{2}(N)$, where $N$ is the number of possible values  $\frac{\phi_1+\phi_2}{\Delta\phi}=\frac{2\phi}{\Delta\phi}$ and $\Delta\phi$ is the search step size. This additional complexity is due the additional search for the phase $\phi^{*}$ that can achieve the minimum transmit power or the highest minimum SNR.  
\section{Symbol Error Rate (SER) Analysis}
\label{SERsection}
In this section, we compare the performance of exact and relaxed detection
in the constructive interference precoding techniques from symbol error
rate (SER) perspective. Assuming any CI precoding technique, the received signal at the $j^{th}$ user
\begin{eqnarray}\nonumber
y_j=(\sqrt{\omega_j}d_j+z_j).
\end{eqnarray}
$\omega_j$ is the received SNR at the $j^{th}$ user. In this section, we drop the index for simplicity. By looking at the received signal and taking its projection on the real and imaginary
axes, the received signal points can be formulated as  
\begin{eqnarray}
y&=&(r_x,r_y)\\\nonumber
&=&(\sqrt{\omega}\cos(\angle d)+\mathcal{R}\{z\},\sqrt{\omega}\sin(\angle d)+\mathcal{I}\{z\})
\end{eqnarray}
where $r_x$, $r_y$ are the projections of the received constellation points on the real
and imaginary axes respectively. Since we assume that $\angle d$ and $\omega$ are fixed, $r_x$, $r_y$  take the distribution of the noise which is independent Gaussian. The corresponding probability density function (PDF) of $r_x,r_y$ can be written as\\
\hspace{-1.2cm}\begin{eqnarray}\nonumber
\label{pdfc}
p(r_x,r_y)&=&\frac{1}{\sigma^2\pi}\exp\Bigg(-\frac{\big(r_x-\sqrt{\omega_{}}\cos(\angle d)\big)^2}{\sigma^2}\Bigg)\\
&\times&\exp\Bigg(-\frac{\big(r_y-\sqrt{\omega_{}}\sin(\angle d)\big)^2}{\sigma^2}\Bigg).
\end{eqnarray}
\normalsize
If we use the polar coordinate transformation $v=\sqrt{r_x^2+r_y^2}$, $\theta=\tan(\frac{r_y}{r_x})$, the previous PDF formulation can be written
as\cite{proakis} 
\begin{eqnarray}\nonumber
\label{pdf_p}
\hspace{-1.2cm}p(v,\theta)=&\frac{v}{\pi\sigma^2}&\exp(-\frac{v^2+\omega+2\sqrt{\omega}v\cos(\angle
d+\theta))}{\sigma^2}\\
&\times&\exp(-\frac{2\sqrt{\omega}v\sin(\angle d+\theta)}{\sigma^2}).
\end{eqnarray}

For the relaxed detection region design, the SER depends on the angular span
$\phi_{j1},\phi_{j2}$. Intuitively, if we increase this span, the received
signal becomes more sensitive to noise. Therefore, the span selection should
depend on the value of SNR. We define a new random variable $\phi_j$ that describes the fact that the transmitted
data symbols can be designed to deviate from the central point of the
detection region. The value of the $\phi$ varies according to the target
SER. Eq. (\ref{pdfc}) can be rewritten to include the impact of relaxation as the following 
 
\begin{eqnarray}\nonumber
\hspace{-0.7cm}p(r_1,r_2,\phi)&=&\frac{1}{\sigma^2\pi}\exp\Bigg(-\frac{(r_1-\sqrt{\omega}\cos(\angle
d+\theta+\phi))^2)}{\sigma^2}\Bigg)\\
&\times&\exp\Bigg(-\frac{(r_2-\sqrt{\omega}\sin(\angle d+\theta+\phi))^2}{\sigma^2}\Bigg).
\end{eqnarray}
\normalsize

Using the polar coordinates, the PDF that describes the
flexible detection region can be expressed as:\\
\normalsize
%\scriptsize
\begin{eqnarray}
\hspace{-0.7cm}p(v,\theta,\phi)=\frac{v}{\pi\sigma^2}\exp\Big(-\frac{v^2+\omega-2v\sqrt{\omega}\cos(\angle
d_k+\theta-\phi)}{\sigma^2}\Big).
\end{eqnarray}
\normalsize
Since we do not consider any SNR target constraint on the system
performance, the SER can be found  by formulating the PDF $p(\theta,\phi)=\int^{\infty}_{0}p(v,\theta,\phi)dv$

\begin{eqnarray}\nonumber
p(\theta,\phi)\hspace{-0.35cm}&=&\hspace{-0.3cm}\exp(-\frac{\sqrt{\omega}\sin(\theta\hspace{-0.08cm}-\hspace{-0.03cm}\phi)}{\sigma^2})\\
&\times&\int^{\infty}_0\hspace{-0.25cm}v\hspace{-0.05cm}\exp(-\frac{(v\hspace{-0.1cm}-\hspace{-0.1cm}\sqrt{\omega}\hspace{-0.08cm}\cos(\theta-\phi))^2}{\sigma^2})dv.
\end{eqnarray}
\normalsize
The generic formulation for the SER for relaxed detection technique can be written as
\begin{eqnarray}
\label{SERr}
P_e&=&1-\int^{\frac{\pi}{M}}_{-\frac{\pi}{M}}\int^{\angle d+\phi}_{\angle d-\phi} p(\theta,\phi)d\theta d\phi.
\end{eqnarray}
For the strict constructive interference precoding techniques, using(\ref{SERr}) can be written
\begin{eqnarray}\nonumber
\label{SERf}
\hspace{-1.9cm}P_e&=&1-\int^{\frac{\pi}{M}}_{-\frac{\pi}{M}}\int^{\angle d}_{\angle d} p(\theta,\phi)d\phi d\theta\\
&=&1-\int^{\frac{\pi}{M}}_{-\frac{\pi}{M}}p(\theta)d\theta.
\end{eqnarray}
The previous formulation is used in \cite{proakis} to derive the SER for any M-PSK modulation.

\section{Trade off analysis}
\label{tradeoff}
%\subsection{SER Analysis with SNR Targets Guarantees}
For the relaxed detection design, the transmitted signals are designed
to be received with controlled deviation from the exact constellation point to enhance the system performance (i.e. minimize
transmit power). The relation between the
improvement achieved by allowing such flexibility and the SER performance
of the system is studied in this section.

We link the SER analysis with CIPM algorithms (\ref{powminr}) in order to find the operating point in terms of phase margin, which minimizes transmit power without considerably degrading the SER.
Using (\ref{SERr}), the SER considering $\zeta_k$ as the minimum acceptable
SNR target can be expressed as 
\begin{eqnarray}
Pe[\omega_k\geq\zeta_k]=\int^{\frac{\pi}{M}}_{-\frac{\pi}{M}}\int^{\angle d_j+\phi}_{\angle d_j- \phi}\int^{\infty}_{\zeta_k}p(v,\theta,\phi)dvd\phi
d\theta.
\end{eqnarray}
The concept of exploiting the relaxed detection gives
the system design
more parameters to be tuned and thus more flexibility and performance gain to be anticipated.
\subsection{Effective Rate Analysis}
The relaxed detection increases the amount of symbol detection errors, which  degrades the rate of each user and affects the performance of whole system. The effective rate for each user can be expressed as  
\begin{eqnarray}
\label{effective_r}
\bar{R}_j\approx R_j\times \big(1-SER (\omega_j,\phi_j)\big)
\end{eqnarray}
$R_{j}$ is the target rate of the employed modulation. 
From (\ref{effective_r}), it can be concluded that enlarging the relaxed detection region increases the SER and as a result decreases the effective rate in the system. 
\subsection{Energy efficiency analysis}
The relaxed detection not only decreases the amount of the power required to achieve the target SNR but also decreases the effective rate of the system. To find the optimal balance between these two aspects, the system energy efficiency metric is proposed to find how many bits can be conveyed correctly to the receivers per energy unit. The system energy efficiency can be defined as 
\begin{eqnarray}
\label{eta}
\eta= \frac{\sum^K_{j=1} \bar{R}_j\Big(SER_j(\omega_j,\phi_j)\Big)}{P(\Phi,\boldsymbol\zeta)}
\end{eqnarray}
where $P(\Phi,\boldsymbol \zeta)=\|\mathbf{x}(\mathbf{H},\mathbf{d},\boldsymbol\zeta,\boldsymbol\Phi)\|^2$. Assuming equal margin, the optimization can be formulated as 
\begin{eqnarray}
\label{energy_efficiency}
\underset{\phi}{\max}\quad\eta
\end{eqnarray}
It should be noted that the energy efficiency is a function of the phase margin since it decreases the transmit power amount required to achieve the target rate. Increasing the phase margin  affects both the numerator and the denominator in (\ref{eta}) by decreasing the effective rate and transmit power respectively. Therefore, the impact of the phase  margin cannot be straightforwardly deduced intuitively. Moreover,
it is hard to solve the optimization problem (\ref{energy_efficiency}) through standard numerical techniques. A simulation-based solution is found in the next section. 
\section{Numerical results}\label{sim}
%\vspace{-0.15cm}

\begin{table}
\begin{center}
\hspace{-0.02cm}\begin{tabular}{|p{1.3cm}|p{5.2cm}|p{1.0cm}|}
\hline
Acronym&Technique&equation\\
\hline
CIZF&Constructive Interference Zero Forcing&\cite{Christos}\\
\hline
CIMRT&Constructive Interference Maximum Ratio Transmissions&\cite{maha}
\\
\hline
CIPM& Constructive Interference- Power Minimization&\ref{CIPM},\cite{maha_TSP}\\
\hline
CIPMR& Constructive interference power minimization with relaxed detection& \ref{CIPMR}\\
\hline
CIMM& Constructive interference max min SNR& \ref{CIMM},\cite{maha_TSP}\\
\hline
CIMMR& Constructive interference maxmin SNR with relaxed detection region& \ref{relaxedp},\textbf{A}\\
\hline
Multicast&Optimal Multicast &\ref{powm1},\cite{multicast}\\
\hline
Genie&Genie theoretical upper-bound&\ref{pr},\cite{maha_TSP}\\
\hline
\end{tabular}
\vspace{0.2cm}
\caption{Summary of the proposed algorithms, their related acronyms, and
their related equations and algorithms}
\end{center}
\end{table}

In order to assess the performance of the proposed transmissions schemes, Monte-Carlo simulations of the different algorithms have been conducted to
study the performance of the proposed techniques and compare to the state
of the art techniques. The adopted channel model is assumed
to be 
\begin{eqnarray}
\mathbf{h}_k\sim\mathcal{CN}(0,\sigma^2).
\end{eqnarray}
$\sigma^2$ is the channel average power.
 
\begin{figure}[h]
\vspace{-0.1cm}
\begin{center}
\vspace{-0.1cm}
\hspace{-0.6cm}\includegraphics[scale=0.58]{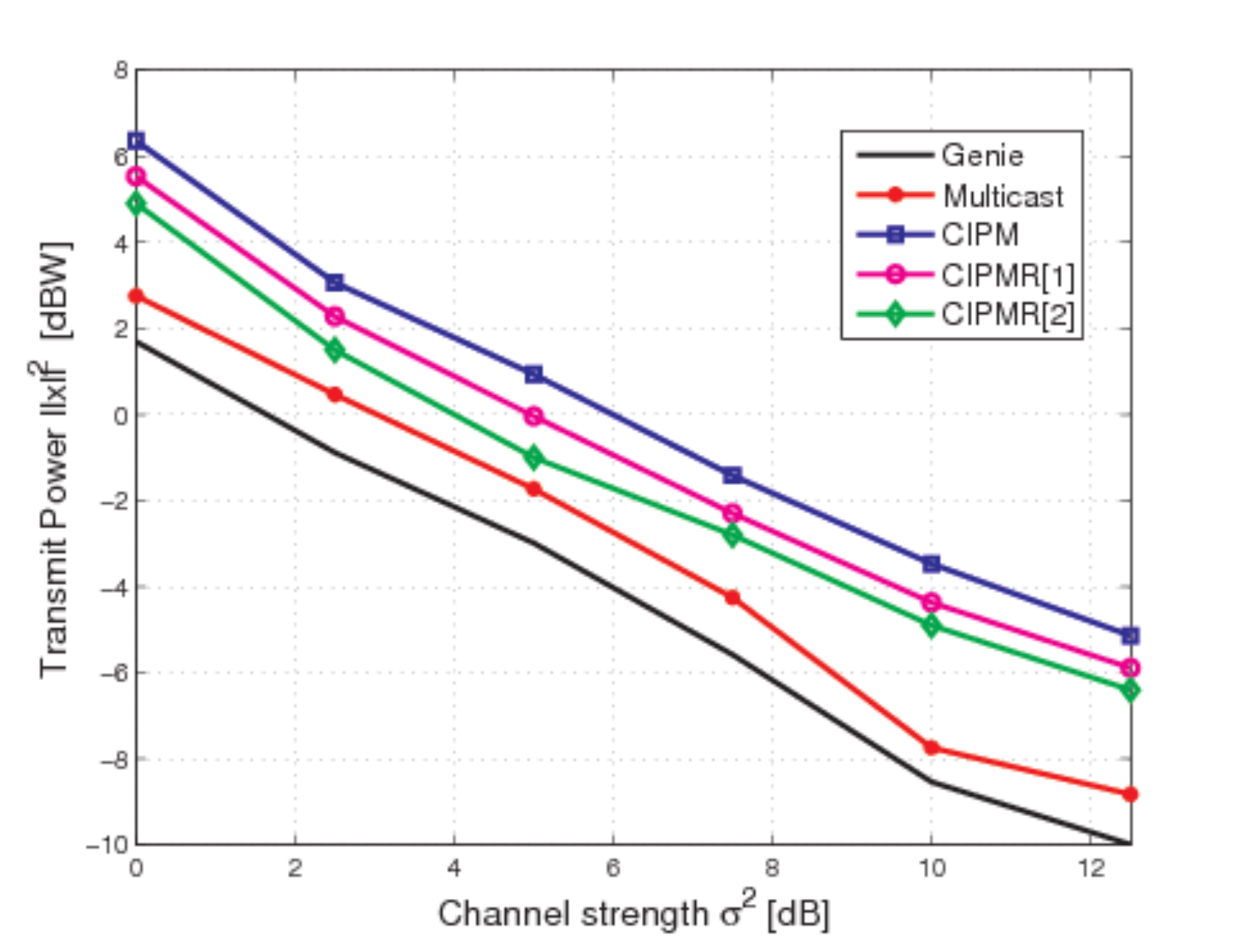}
\caption{\label{po}Transmit power $\|\mathbf{x}\|^2$ vs channel strength $\sigma^2$. CIPMR[1] denotes the scenario of $\phi=\frac{\pi}{8}$ and CIPMR[2] denotes the scenario of $\phi=\frac{\pi}{5}$.$M=3$, $K=2$, $\zeta= 4.7712 dB$, QPSK.}
\end{center}
\end{figure}
Fig. (\ref{po}) depicts the power consumption with respect to target SNR.
The comparison among optimal multicast, CIPM and CIPMR is illustrated in this figure while the assumed scenario is $M=3$, $K=2$, at $\phi=\frac{\pi}{5}$ and $\frac{\pi}{8}$. It can concluded that the power consumption gap between the optimal multicast and CIPM is fixed for all target rates. This relation holds also for the gap between the CIPMR and CIPM. Moreover, it can be concluded that the CIPMR outperforms CIPM by achieving less power at all target SINR values. Moreover, CIPMR
 shows a better performance at $\phi=\frac{\pi}{5}$  than $\phi=\frac{\pi}{8}$.
 
\begin{figure}
\begin{center}
\hspace{-0.6cm}
\hspace{-0.6cm}\includegraphics[scale=0.6]{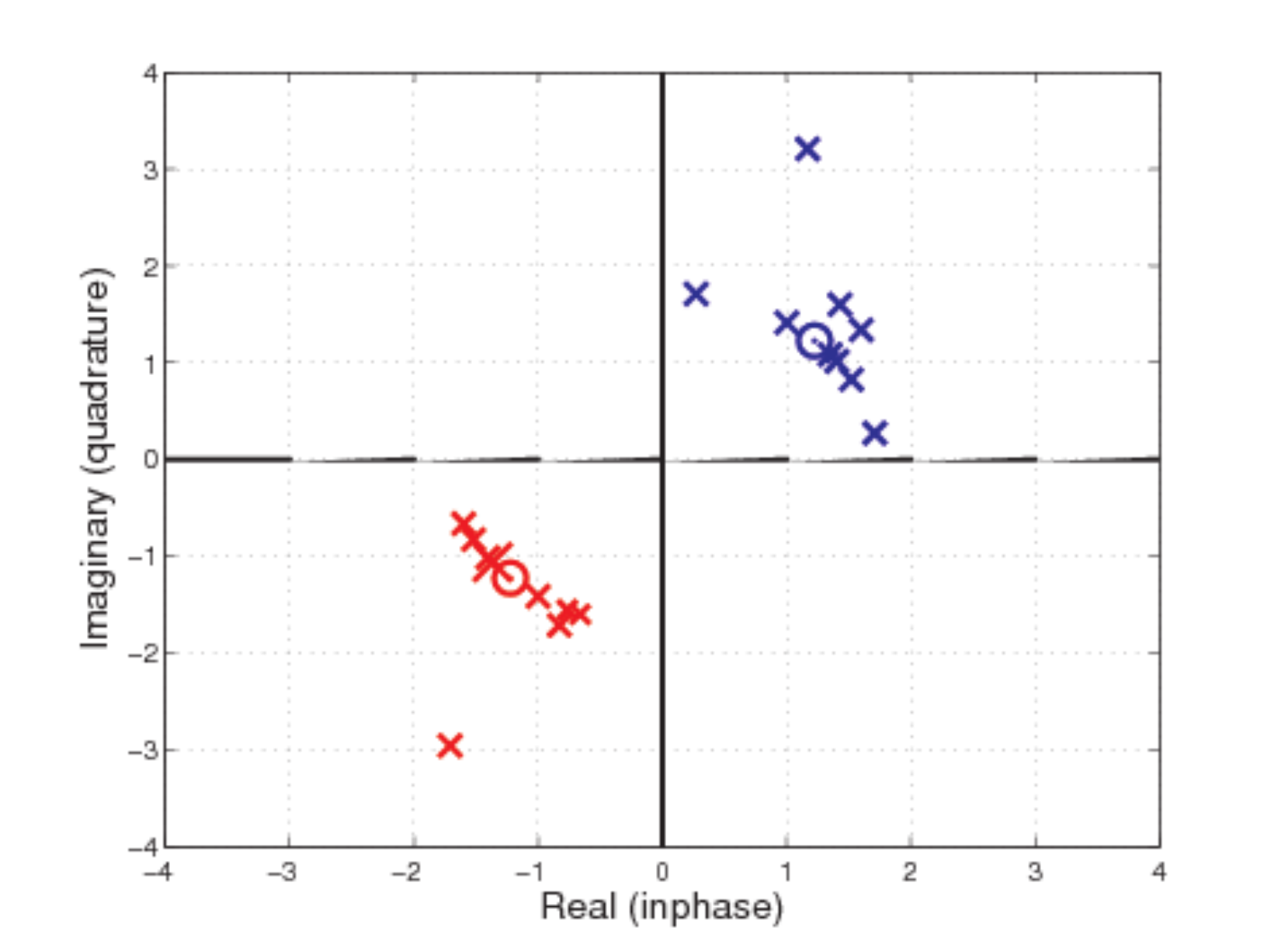}
\caption{\label{inphase} The received signal using CIPMR and CIPM without the noise effect. $M=3$, $K=2$, $\phi=\frac{\pi}{5}$, $\zeta=4.7121dB$, $\sigma^2=0dB$, QPSK. The circles denote the detected signal at the receivers assuming CIPM, the crosses denote the detected signal at the receivers assuming CIPMR. }
\end{center}
\end{figure} 

Fig. (\ref{inphase}) depicts the detected signals at users' receivers. The first user should receive $``11"$, which should be detected at the first quadrant. The second user should receive the symbol $``00"$, which should be detected at the third quadrant. The number of the transmitted symbols for each user is $``10"$ symbols. It can be noted that the received signals using CIPMR has higher power than the target SNR. In these cases, the received power at the first user is equal to the target SNR while the other detects its symbol with power higher than its target even though less power is actually used for transmission.   This means that the algorithm searches for the phases of the data symbols  in the relaxed region that grants the minimum power; sending with certain phase aids the other user and pushes the symbol deeper in the detection region.    

\begin{figure}[h] 
\begin{center}
\vspace{-0.1cm}
\hspace{-0.6cm}\includegraphics[scale=0.65]{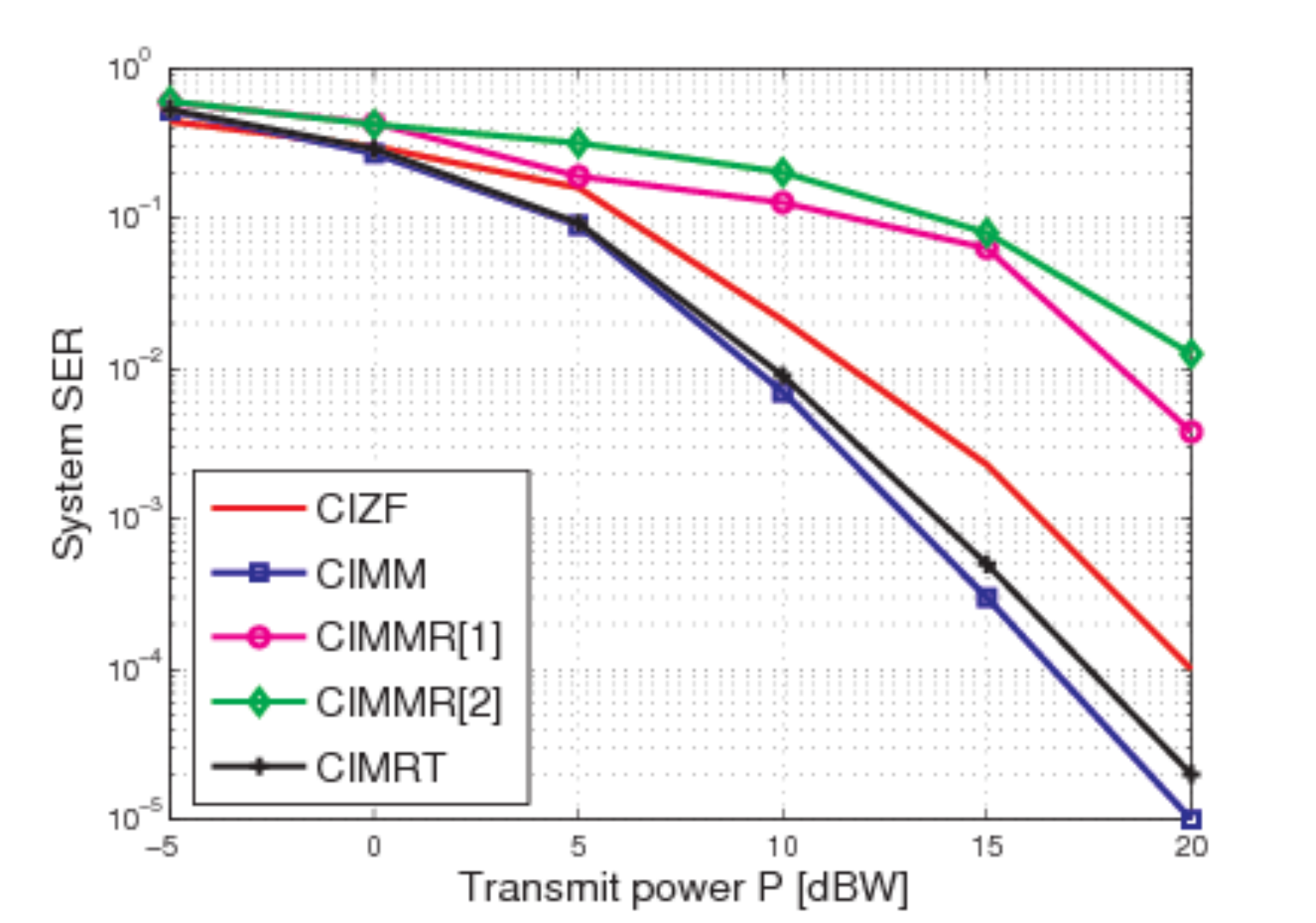}
\caption{\label{SER} SER vs transmit power. CIMMR[1] denotes the scenario of $\phi=\frac{\pi}{8}$ and CIMMR[2] denotes the scenario of $\phi=\frac{\pi}{5}$. $M=3$, $K=2$, $\sigma^2=0dB$, QPSK.}
\end{center}
\end{figure}

The system SER performance with respect to the available power is depicted in Fig.(\ref{SER}). It can be noted that CIMM has the lowest SER. At 20 dB, the SER of CIMM is around $10^{-5}$ without employing any FEC coding. Moreover, it can be noted that CIMRT has very close performance to CIMM. It also can be deduced that CIZF has a higher SER than CIMM and CIMRT across the studied power range. As expected, CIMMR has the worst performance in comparison with the other techniques. Varying the angular span of relaxation affects the SER, the SER in the scenario of $\phi=\frac{\pi}{5}$ is close to $10^{-2}$ at 20 $dB$, and the SER in the scenario of $\phi=\frac{\pi}{8}$ is around $5\times10^{-3}$, which is almost half of the value at the scenario of $\phi=\frac{\pi}{8}$. 
 
\begin{figure}[h]
\begin{center}
\vspace{-0.1cm}
\includegraphics[scale=0.6]{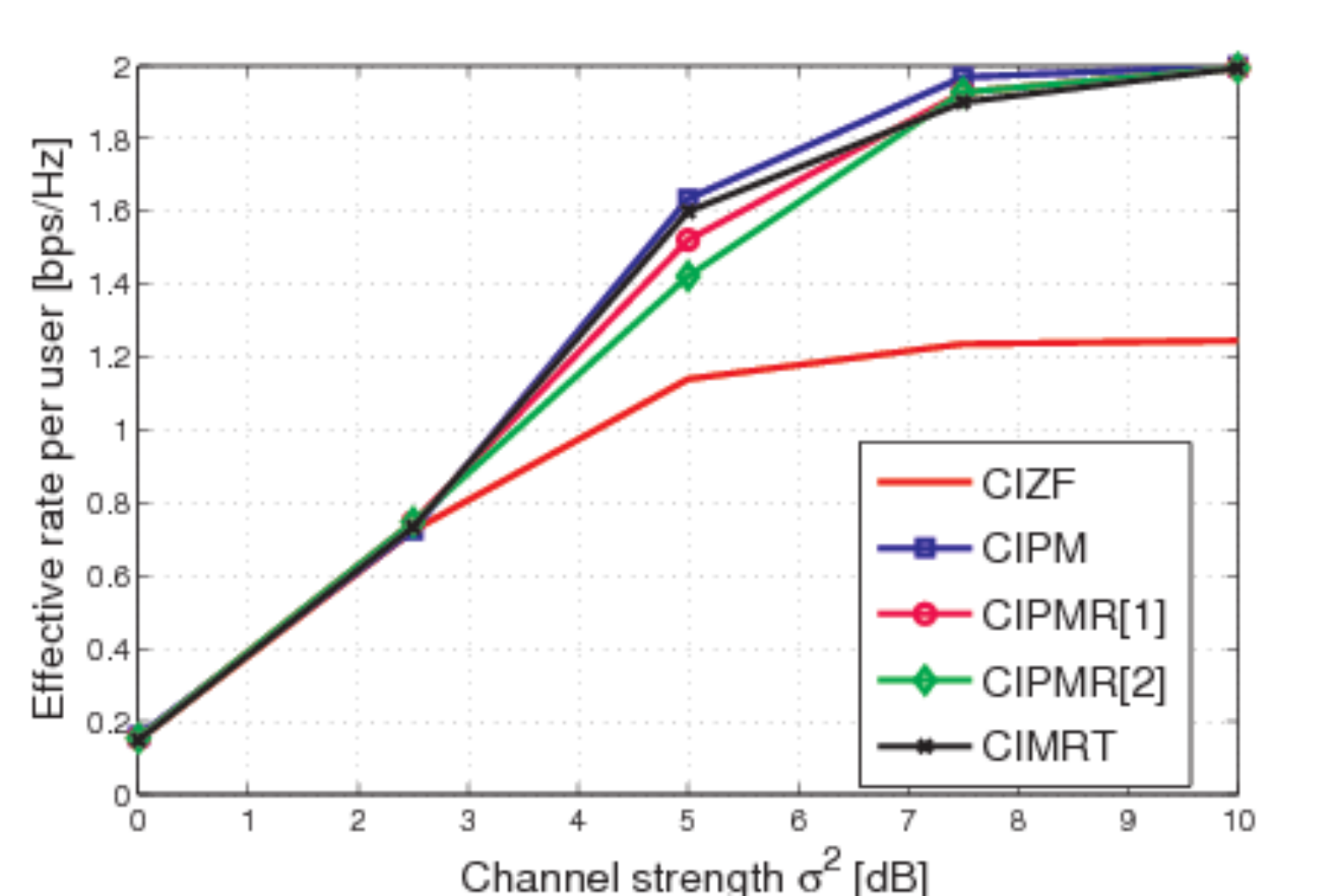}
\caption{\label{rate_qpsk}Rate per user vs. channel strength. CIPMR[1] denotes $\phi=\frac{\pi}{8}$ and CIPMR[2] denotes the scenario of $\frac{\pi}{5}$. $M=3$, $K=2$, $\zeta=4.7121 dB$, QPSK.}
\end{center}
\end{figure}

The effective rate per user versus the channel strength is depicted in Fig. (\ref{rate_qpsk}). The general trend is that the rate increases with the available power. However, the slope of each curve indicates the amount of rate increase with respect to the available power. Although it has a reasonable SER performance, it can be noted that CIZF has the worst performance from the rate perspective. On the other hand, CIPM achieves the best performance since it has the lowest SER values. CIMRT has a very close performance to CIPM. Regarding the relaxed detection region approach, at $\phi=\frac{\pi}{8}$, the system has a better rate performance than at the scenario $\phi=\frac{\pi}{5}$ due to the higher SER at the latter case. Moreover, all the techniques perform the same at low SNR.

 \begin{figure}[h]
\begin{center}
\hspace{-0.8cm}\includegraphics[scale=0.6]{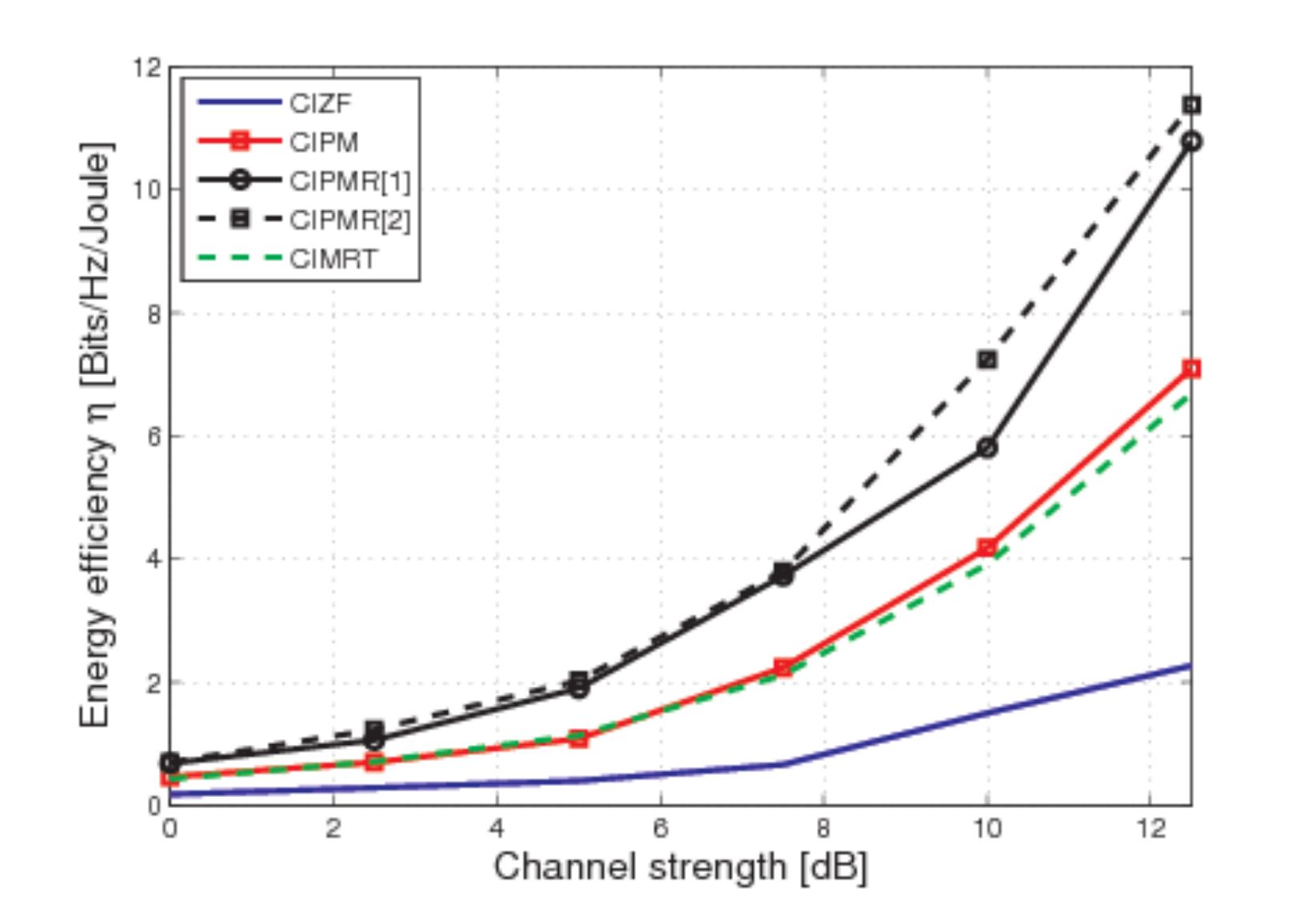}
\caption{\label{energyefficiency}Energy efficiency $\eta$ vs channel strength $\sigma^2$.  CIPMR[1] denotes the scenario of $\phi=\frac{\pi}{8}$ and CIMPR[2] denotes the scenario of $\phi=\frac{\pi}{5}$. $M=3$, $K=2$, $\sigma^2=0dB$, QPSK.}
\end{center}
\end{figure}
\begin{figure}[h]
\begin{center}
\vspace{-0.1cm}
\hspace{-0.2cm}\includegraphics[scale=0.64]{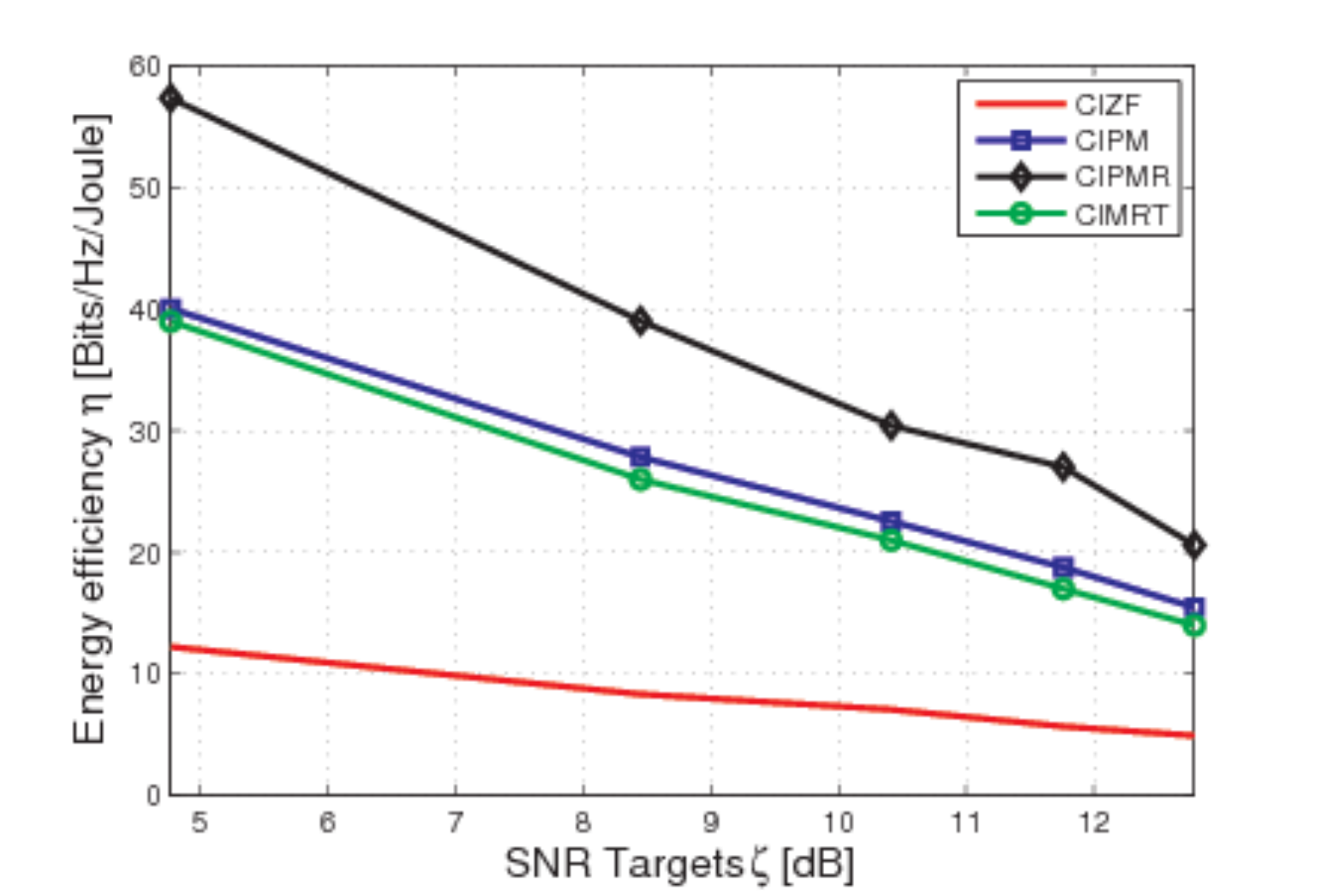}
\caption{\label{eevs}Energy efficiency $\eta$ versus SNR targets $\zeta$. $M=3$, $K=2$, QPSK, $\sigma^2=20 dB$, $\phi=\frac{\pi}{5}$.}
\end{center}
\end{figure}
In Fig. (\ref{energyefficiency}), we depicted the performance of the proposed techniques from energy efficiency
perspective with the channel strength. CIZF shows inferior performance in comparison with all
depicted techniques. It has already been proven that CIZF 
outperforms the conventional techniques like minimum mean square error (MMSE)
beamforming and zero forcing beamforming (ZFB) \cite{Christos}. In comparison with other depicted techniques, it can be concluded that
the proposed constructive interference CIPM and CIPMR have better energy efficiency in comparison with CIZF.
This can be explained by the channel inversion step in CIZF which wastes energy in
decoupling the effective users' channels and before exploiting the
interference among the multiuser streams. Moreover, it can be noted that the CIPMR achieves higher energy efficiency than CIPM, since it allows selecting flexibly the target point inside the detection region.  Moreover, it can be deduced that
CIMRT has a very close performance to CIPM especially at high targets. CIMRT
outperforms CIZF at expense of complexity.

\begin{figure}[h]
\begin{center}
%\vspace{-0.1cm}
\hspace{-0.2cm}\includegraphics[scale=0.555]{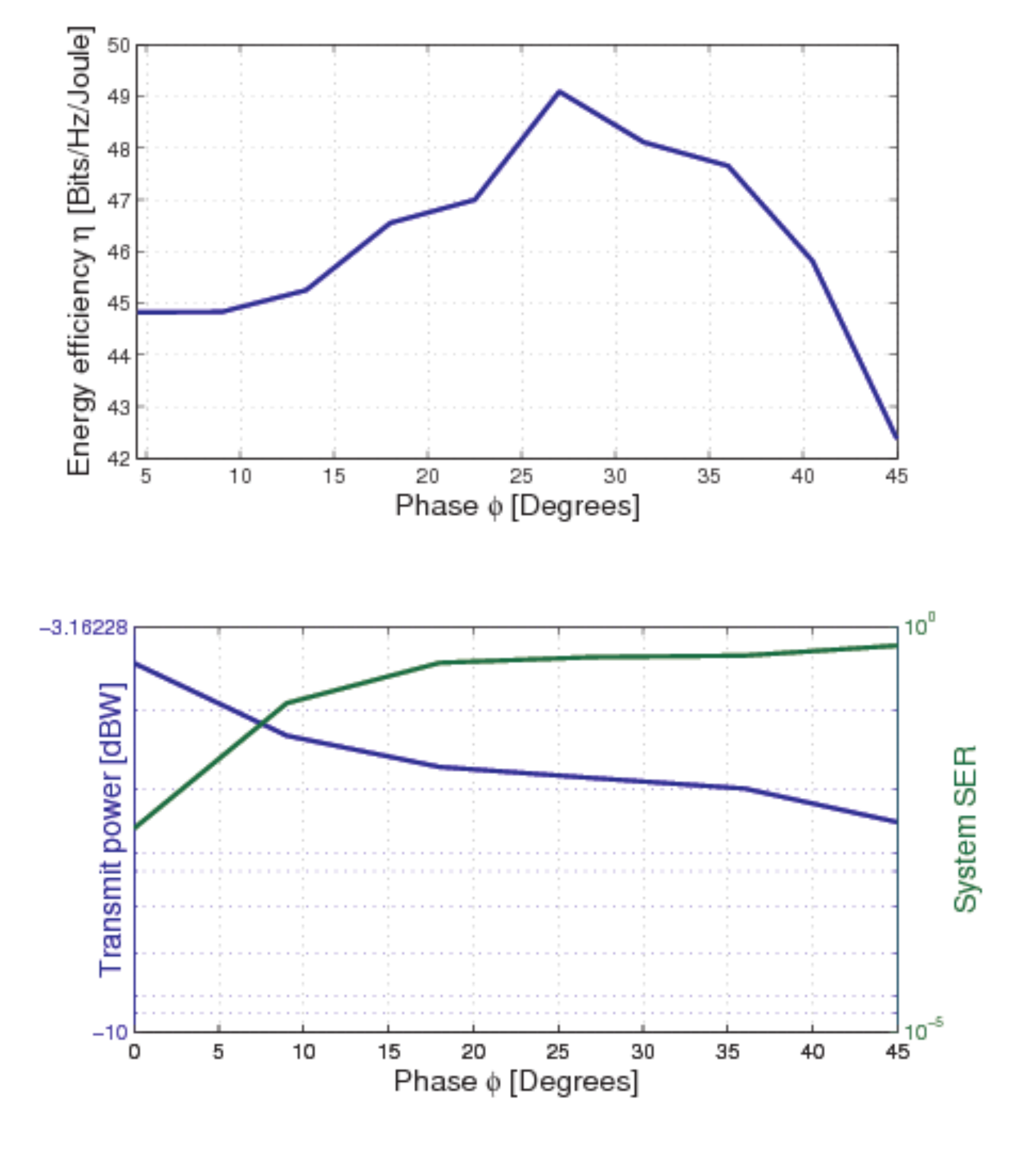}
\caption{\label{ds_ser} Energy efficiency and SER vs angular span $\phi$. $M=3$, $k=2$, QPSK, $\zeta=13.01 dB$, $\sigma^2=20dB$.}
\end{center}
\end{figure}

Fig. (\ref{eevs}) depicts the energy efficiency with respect to SNR targets. We depict the performance of CIZF and CIMRT.
For the sake of comparison, the transmit power of the CIZF and CIMRT solutions can be scaled until all users achieve the target rate.\smallskip It can be noted that CIPMR has a better performance in comparison with the other techniques. The CIPMR has a higher gap at low target SNR values.  

Furthermore, the flexibility should be adapted with the target rates. At low target SNR, the flexibility region should be narrowed to prevent from moving outside the correct detection region due to noise. At high target SNR, the flexibility region can be enlarged since the impact of noise can be negligible. This impact is depicted in Fig.(\ref{ds_ser}) at high target SNR. Increasing the angular span of the relaxation decreases the transmit power and increases SER. The two factors influence the energy efficiency of system. Increasing the angular span enhances the energy efficiency to certain limit 
$\phi=27^{\circ}$, and start decreasing gradually, Moreover, it can be noted that SER increases with increasing the angular span.\\
\begin{figure}[h]
\begin{center}
\vspace{-0.1cm}
\includegraphics[scale=0.66]{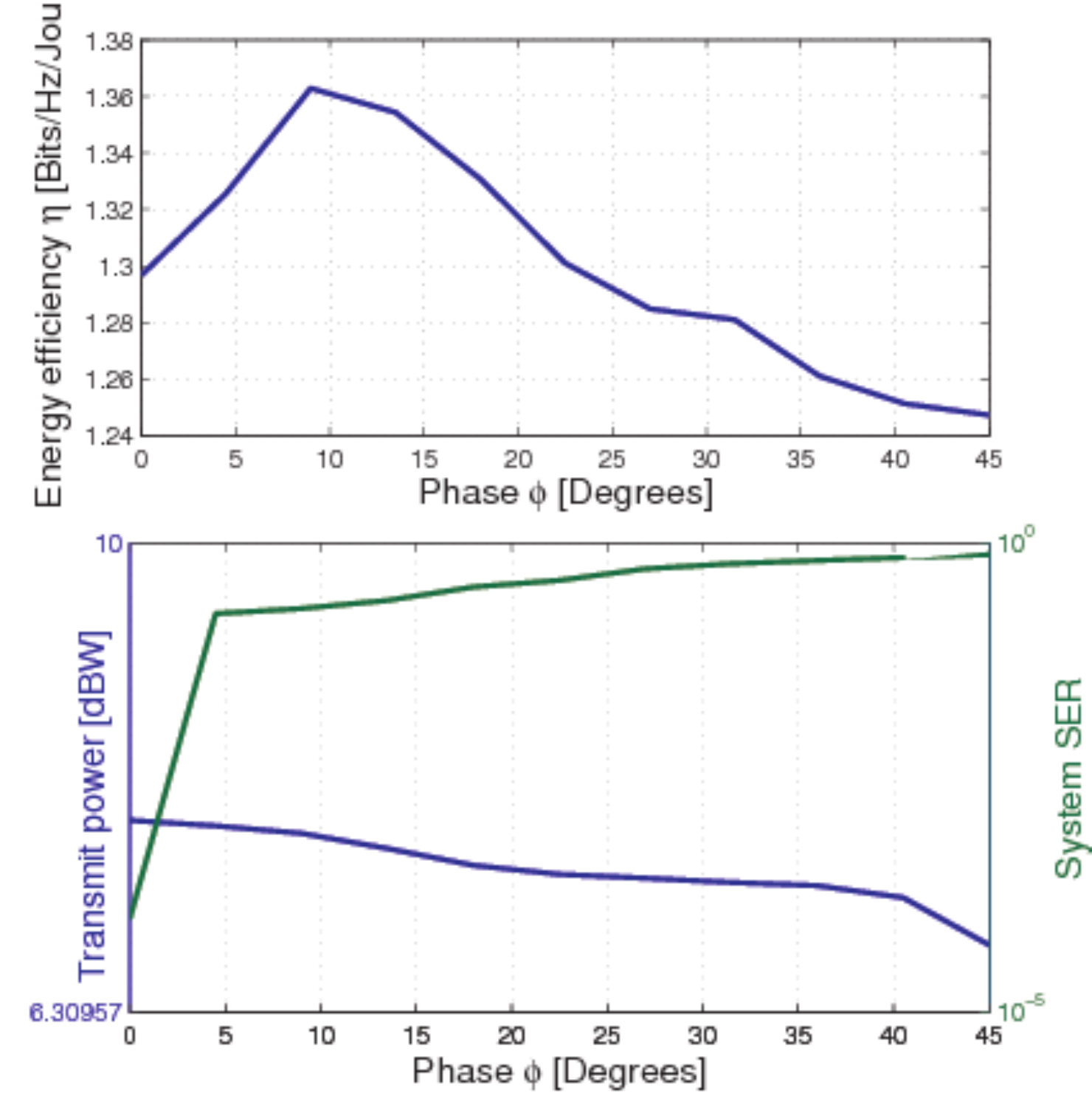}
\caption{\label{DS_SER_LF}Energy efficiency and SER vs angular span $\phi$. $M=3$, $K=2$, $\sigma^2=0 dB$, $\zeta= 4.7712 db$, QPSK.}
\end{center}
\end{figure}
 The effect of the flexible angular span at a low SNR target scenario is depicted in fig. (\ref{DS_SER_LF}). It can be noted that the highest energy efficiency is achieved at $\phi=10^{\circ}$, which is much lower than the $\phi$ value that achieves the highest energy efficiency at 20 $dB$. This means that the impact of the symbol errors overcomes the transmit power saving at a much narrower phase margin due to the low SNR. Hence, the result confirms that the optimal phase margin is a function of the SNR targets.

\begin{table}
\begin{center}
\hspace{-0.02cm}\begin{tabular}{|p{1.7cm}|p{1.2cm}|p{1.2cm}|p{1.2cm}|p{1.2cm}|}
\hline
Mod|$\phi$& 0 &$\frac{\pi}{16}$&$\frac{\pi}{8}$&$\frac{3\pi}{16}$\\
\hline
BPSK&67.75&69.9&71&72.5\\
\hline
QPSK&135.5&139.1&136&121\\
\hline
\end{tabular}
\vspace{0.2cm}
\caption{\label{comparison} Comparison between BPSK and QPSK from energy efficiency perspective. $M=3$, $k=2$, $\sigma^2=20dB$, $\zeta=4.712dB$. }
\end{center}
\end{table}

Table (\ref{comparison}) illustrates the comparison between QPSK and BPSK modulations in terms of energy efficiency assuming CIPMR. We use the same SNR target for the both modulations. It can be concluded that the energy efficiency increases with the relaxation due to the larger angular span of the BPSK detection region. However, QPSK has a different trend; the energy efficiency increases with the relaxation phase up to a point, and decreases afterwards. This trend is expected to occur at a higher phase in BPSK.    
\section{conclusions}
In this paper, we design an energy efficient precoding in the downlink  
of a MU-MISO system. The main idea is based on exploiting the interference among the multiuser transmissions while using symbol-based precoding in combination with MPSK modulations. Particularly, we utilize the concept that the detection region of an M-PSK symbol spans a range of phases, which enables us to relax the system design and to achieve higher power savings. This can be implemented by allowing the precoder to select the optimal phase for each user symbols that can achieve the minimum power without being erroneously detected at the receiver. However, such relaxation increases the system SER. The trade off between the achieved power saving and the SER is characterized by the energy efficiency. The phase margin of the relaxed region can be optimally selected to achieve the highest energy efficiency. The simulation results have confirmed that the relaxed system designs achieve higher energy efficiency especially in the high SNR regime.

\end{document}